\documentclass[twocolumn,english,reprint,prb,showpacs,superscriptaddress]{revtex4-1}
\usepackage{epsfig}
\usepackage{graphicx}
\usepackage{amsfonts}
\usepackage[figuresright]{rotating}
\usepackage{amssymb}
\usepackage{amsmath}
\usepackage{psfrag}
\usepackage{amsmath}
\usepackage{graphicx}
\usepackage{color}

\usepackage{verbatim}
\usepackage{hyperref}

\begin{document}

\title{Interaction effects on topological phase transitions via numerically exact quantum Monte Carlo calculations}
\author{Hsiang-Hsuan Hung}
\affiliation{Department of Physics, The University of Texas at Austin, Austin, Texas, 78712, USA }

\author{Victor Chua}
\affiliation{Department of Physics, The University of Texas at Austin, Austin, Texas, 78712, USA }
\affiliation{Department of Physics, The University of Illinois at Urbana-Champaign, Illinois, 61801, USA }

\author{Lei Wang}
\affiliation{Theoretische Physik, ETH Zurich, 8093 Zurich, Switzerland}

\author{Gregory A. Fiete}
\affiliation{Department of Physics, The University of Texas at Austin, Austin, Texas, 78712, USA }

\begin{abstract}
We theoretically study topological phase transitions in four
generalized versions of the Kane-Mele-Hubbard model with up to $2\times 18^2$ sites.
All models are free of the fermion-sign problem allowing
numerically exact quantum Monte Carlo (QMC) calculations to be performed
to extremely low temperatures.  We numerically compute the $\mathbb{Z}_2$
invariant and spin Chern number $C_\sigma$ directly from the zero-frequency
single-particle Green's functions, and study the topological phase transitions driven by the tight-binding parameters at different on-site interaction strengths. The $\mathbb{Z}_2$ invariant and spin Chern number, which are complementary to each another, characterize the topological phases and identify the critical points of topological phase transitions.  Although the numerically determined phase boundaries are nearly identical for different system sizes, we find strong
system-size dependence of the spin Chern number,  where quantized values are only expected upon approaching the thermodynamic limit.   For the Hubbard models we considered, the QMC results show that correlation effects lead to shifts in the phase boundaries relative to those in the non-interacting limit, without any spontaneously symmetry breaking. The interaction-induced shift is non-perturbative in the interactions and cannot be captured within a ``simple" self-consistent calculation either, such as Hartree-Fock.  Furthermore, our QMC calculations suggest that quantum fluctuations from interactions stabilize topological phases in systems where the one-body terms preserve the $D_3$ symmetry of the lattice, and destabilize topological phases when the one-body terms  break the $D_3$ symmetry.

\end{abstract}

%\date{\today}

\maketitle

\section{Introduction}

Electron interactions in topological insulators\cite{Young:prb08,Pesin:np10,Kargarian:prb10,Li:np10,hohenadler2011,hohenadler2012,zheng2011,Swingle:prb11,Maciejko:prl10,Kargarian:prl13,Levin:prb12,Levin:prl09,Moore:prl08,Neupert:prb11,Ruegg:prl12,Wan:prb11,Hohenadler:jpc2012,Fiete:pe11,Maciejko:prl14,Maciejko:prb13}
have been a topic of intense research in recent years.\cite{Moore:nat10,Hasan:rmp10,Qi:rmp11,Konig:sci07,Roth:sci09,Hsieh:nat08}
It is crucial to go beyond the mean-field
level\cite{Raghu:prl08,Zhang:prb09,Wen:prb10,Liu:prb10,Ruegg11_2} to
capture important fluctuation effects originating in the electronic
correlations, as this can be decisive in determining the
phase.\cite{Go:prl12,hung2013,Budich:prb12,Budich:prb13,Wang:epl12,Yoshida:prb13}
Exact diagonalization studies\cite{Varney:prb10} are
inhibited by significant finite-size effects,\cite{Varney:prb11} though they are unbiased by any particular ansatz in the way mean-field theories are. In
this work, we study correlation effects in topological insulators by considering the Kane-Mele-Hubbard model and several variants with numerically exact
projective quantum Monte Carlo (QMC) calculations.  Due to their particle-hole symmetry, these models are free of the fermion minus sign problem, and QMC simulations provide a great opportunity to study correlation effects in topological matter with an unbiased theoretical approach. We are able to accurately treat correlations\cite{hung2013,lang2013}, compute interacting topological invariants such as the $\mathbb{Z}_2$ invariant and spin Chern number, and identify topological phase transitions through the zero-frequency single-particle Green's
function.\cite{Wang:prx12,Wang:prb12}  We thus avoid complications
associated with ground-state evolution under twisted boundary
conditions,\cite{Niu:prb85} where numerical computations of a manifold ground states is required, and 
potential subtleties regarding energy gap closures must be addressed.\cite{Fukui:prb07} 

Strictly speaking, the ground state is altered when twisted boundary conditions are introduced. The existence of a family of ground states smoothly connected to one and another, and a  finite spectral gap are required for the use of twisted boundary conditions. Meeting these conditions can be especially challenging when approaching a phase transition where excitation gaps can become very small.  Moreover, the use of the twisted boundary conditions is not practical in large-scale simulations, such as QMC;  hence current implementations of twisted boundary conditions in interacting models have mainly focused on small sizes where exact diagonalization techniques have been used.\cite{Varney:prb10,Varney:prb11}  In addition, the initial use of twisted boundary conditions for defining the spin Chern number introduced edge effects, which initially cast doubt on its robustness as a bulk topological invariant.\cite{Fukui:prb07,fu2007,prodan2009prb}

Therefore, it is worth revisiting topological phase transitions from the point-of-view of spin Chern numbers, particularly in the context of systems with interactions and finite-size effects present.  We observe in our numerical QMC results a dichotomy in the role of on-site Hubbard interactions: Depending on the underlying lattice symmetry, they favor either a topological or trivial phase. Although our results are only limited to the class of models with particle-hole symmetry and $S^z$ conservation, they could be hints of a more general principle regarding the interplay between point-group symmetry and interactions. 

The remainder of this paper is organized as follows. In Section \ref{sec:KM}, we introduce the Kane-Mele model and the four variants of it that we study. We compare and contrast the particular spatial symmetries exhibited by these toy models. In Section \ref{sec:top_ind}, we follow up with a discussion on the time-reversal invariant topological $\mathbb{Z}_2$ index and the spin Chern number with a focus on their numerical implementation in the presence of interactions. Next in Section \ref{sec:int} which is the main part of our work, we present computations of topological indices in the presence intermediately strong interactions for the models introduced in Section \ref{sec:KM}. This is followed up with discussions, interpretation, and speculation regarding these results in Section \ref{sec:discussion}. Then in Section \ref{sec:summary} we conclude with a summary and conclusions. Also included in the appendices are details regarding our quantum Monte Carlo methodology and supporting numerical results on the spin Chern number.

%%%%%%%%%%%%%%%%%%%%%%%%%%%%%%%%%%%%%%%%%%%%%%%%%%%%%%%%%%%%%%%%%%%%%%%%%%%%%%%%%%%%%%

\section{The Kane-Mele Model and several variants}
\label{sec:KM}

\begin{figure}[!t]
\epsfig{file=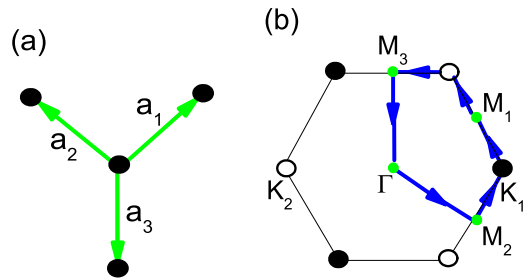,clip=0.1,width=0.8\linewidth,angle=0}
\caption{(Color online) (a) The honeycomb lattice and the underlying vectors ${\bf a}_{1,2,3}$. (b) The first Brillouin zone of the
honeycomb lattice. The Dirac points are
$K_{1,2}=(\pm\frac{4\pi}{3\sqrt{3}a},0)$ labeled by open and solid
circles, respectively. The time-reversal invariant momentum (TRIM)
points labeled by the green dots are $\Gamma=(0,0)$ and
$M_{1,2}=(\pm\frac{\pi}{\sqrt{3}a},\frac{\pi}{3a})$ and
$M_{3}=(0,\frac{2\pi}{3a})$.} \label{fig:KMmodel}
\end{figure}

The Kane-Mele (KM) model, an early model supporting a $\mathbb{Z}_2$ topological
insulator (TI) on the honeycomb lattice,\cite{ kane2005a,kane2005b}
remains central to the study of interaction effects in TI. The
honeycomb lattice is a Bravais (triangular) lattice with a two-point
basis (labeled as $A$ and $B$). The vectors connecting two
neighboring sites are ${\bf
a_{1,2}}=\pm\frac{\sqrt{3}}{2}a\hat{x}+\frac{1}{2}a\hat{y}$ and
${\bf a_{3}}=-a\hat{y}$, where $a$ is the lattice constant between
two nearest-neighbor sites as shown in Fig. \ref{fig:KMmodel} (a); we set $a=1$ hereafter.  The Hamiltonian reads as
\begin{equation}
H_{KM}  =  - t\sum_{\langle
i,j\rangle}\sum_{\sigma}c_{i\sigma}^{\dagger}c_{j\sigma}+i
\lambda_{SO}\sum_{\langle\langle i,j\rangle\rangle}
\sum_{\sigma}\sigma c^{\dagger}_{i\sigma}\nu_{ij} c_{j\sigma}, \nonumber 
\label{eqn:dimerham}
\end{equation}
where $c^{\dag}_{i\sigma} (c_{i\sigma})$ creates (annihilates) a
spin $\sigma$ fermion on site $i$ and $\sigma$ runs over $\uparrow$
and $\downarrow$. Here, $\langle\langle \cdots \rangle\rangle$
denotes second-neighbor terms given by vectors ${\bf
b_{1}}={\bf a_2}-{\bf a_3}$, ${\bf b_{2}}={\bf a_3}-{\bf a_1}$ and
${\bf b_{3}}={\bf a_1}-{\bf a_2}$ describing the spin-orbit coupling $\lambda_{SO}$.
$i=1,2,3$, and $ \nu_{ij}=1$ for counter-clockwise hopping and
$\nu_{ij}=1$ otherwise.\cite{ kane2005a}

For our numerical study, we consider {\em four} time-reversal symmetric model Hamiltonians which are KM model-like:
(i) the generalized Kane-Mele (GKM) model,  the KM model with a
spin-independent real-valued third-neighbor hopping term, (ii) the dimerized Kane-Mele (DKM), the KM
model with a biased nearest-neighbor hopping along the ${\bf
a_1}$ direction, (iii) the $t_L$-KM model, the KM model with $5$-th neighbor  hopping, and (iv) the $t_{3N}$-dimerized KM model, which is 
a hybrid of model (i) and (ii).

 All of the models are generalized versions of the KM models, and at half-filling, they
preserve the particle-hole symmetry. In the non-interacting limit, 
they host a topological-insulator/trivial insulator phase transition by tuning tight-binding parameters.
However, there exists crucial differences among the models.
The GKM model (i) and the $t_L$-KM model (iii) preserve the six-fold rotation or $C_6$ symmetry of the 
honeycomb lattice, whereas the DKM model (ii) and the $t_{3N}$-dimerized model (iv) explicitly break it down to $C_2$ with 
the bias in the ${\bf a_1}$ direction. In the following we 
will mainly focus on the GKM and DKM models.

The wallpaper or 2D space group of the honeycomb lattice is {\bf p6m} which is \emph{symmorphic} and has $D_6$ as its point group. The different models considered are meant to represent different modifications of the bare KM model such that the $D_6$ symmetry is either preserved or broken but which nevertheless exhibits a topological insulator phase transition. However, time-reversal, spin-$S^z$, inversion, and particle hole symmetry are preserved in all these models. More precisely these models exhibit the Quantum Spin Hall (QSH) phase when topologically non-trivial under the $\mathbb{Z}_2$ classification of 2D time-reversal symmetric topological insulators. It is well known that the KM model includes spin-orbit coupling in the form of spin-dependent second nearest neighbor hopping that favors the topological insulator phase. We shall see in models (i-iv) that hoppings which are additions to the standard KM model will, at sufficient strengths, overcome this tendency of the spin-orbit coupling and stabilize the trivial phase without breaking any symmetries. 

Resuming our discussion on symmetry, recall that $D_6 \cong D_3 \times \mathbb{Z}_2^{(i)}$, where the $(i)$ superscript in $\mathbb{Z}_2^{(i)}$ denotes 2D inversion about the center of the hexagon. Furthermore $D_3 \cong C_3 \rtimes \mathbb{Z}_2^{(m)}$ where the $(m)$ denotes reflection about a vertical mirror plane i.e. $\mathbb{Z}_2^{(m)} \equiv \sigma_v$ in Sch\"onflies notation. Moreover, in two dimensions inversion is isomorphic to rotation by 180$^\circ$ or $\mathbb{Z}_2^{(i)} \cong C_2$. Models (i-iv) are selected to maintain $\mathbb{Z}_2^{(i)}$ inversion symmetry but may either preserve or break $D_3$ down to $\mathbb{Z}_2^{(m)}$ or completely. A further essential property that is common to these models is the absence of QMC sign problems in their respective Hubbard model incarnations, which are obtained by the inclusion of an on-site Hubbard interaction. 

Specializing to two dimensions,  the seminal works of Kane and Mele\cite{kane2005a,kane2005b}, and later  Bernevig {\it et. al.}\cite{bernevig2006}, Schnyder {\it et. al}.\cite{schnyder2009classification}, Kitaev,\cite{kitaev2009periodic} and Qi {\it et. al.}\cite{qi2008} showed that with only time-reversal symmetry the non-interacting topological phases are classified by a $\mathbb{Z}_2$ invariant, which is also generalized to three dimensions by Fu and Kane\cite{ fu2007prl}, and Moore and Balents\cite{ moore2007}. The physical content of this binary topological index is that it enumerates the number parity of Kramers pairs of gapless edge modes at a boundary of the system with the vacuum--at least for non-interacting gapped band insulators. With the absence of $S^z$ mixing, the non-interacting occupied bands may be further categorized by their $S^z$ polarization, and each spin species is topologically non-trivial carrying a non-zero integral Chern number; {\it i.e}. $C_\sigma\neq 0, \; \sigma = \uparrow,\downarrow$.\footnote{We have also specialized to the case where there is also only a single valence and conduction band per-spin species.} However, due to time-reversal symmetry, $C_{\uparrow}+C_{\downarrow}=0$, and hence we do not expect an Integer Quantum Hall effect. Nevertheless, if the spin Chern number\cite{sheng2006,shengli2005,prodan2009prb} defined by
\begin{align}
C_\text{spin}=\frac{C_\uparrow-C_\downarrow}{2} = C_{\uparrow} = -C_\downarrow
\label{eqn:ChernSpin}\end{align}
is odd and non-zero, then the ground state of filled bands are in the non-trivial topological insulator phase or the $\mathbb{Z}_2$ odd phase. The non-mixing of $S^z$ sectors, designates this non-trivial phase as being the QSH phase where on the edge, an odd number of helical edge states carrying $S^z$-current persists so long as time-reversal symmetry and the bulk band gap are preserved. As was stated, all the models we have considered will either be trivial or in the QSH phase in the non-interacting limits.  In this work, we will  demonstrate that the classification by $C_\text{spin}$ will not only be applicable in the non-interacting limit, but can also be extended to finite interaction where our main interests lie. It is also clear that under this classification, a topological phase transition between even and odd values of $C_\text{spin}$ must proceed by an odd variation $\Delta C_\text{spin} \in 2\mathbb{Z}+1$. 

It will be highlighted in the upcoming sections that crystal symmetry will play a crucial role in the nature of such topological transitions.

\subsection*{(i) Generalized Kane-Mele model}

We start with the GKM model previously introduced in Ref.[\onlinecite{hung2013}] whose Hamiltonian is given by 
\begin{eqnarray}
 H_{GKM} & = &
- t\sum_{\langle
i,j\rangle}\sum_{\sigma}c_{i\sigma}^{\dagger}c_{j\sigma}+i
\lambda_{SO}\sum_{\langle\langle i,j\rangle \rangle}
\sum_{\sigma}\sigma c^{\dagger}_{i\sigma}\nu_{ij} c_{j\sigma}
\nonumber \\  & &  - t_{3N}\sum_{\langle\langle\langle
i,j\rangle\rangle\rangle}\sum_{\sigma}c_{i\sigma}^{\dagger}c_{j\sigma},
\label{eqn:GKMham}
\end{eqnarray}
where $\langle\langle\langle \cdots
\rangle\rangle\rangle$ third-neighbor terms, and the vectors connecting
third-neighbor terms are given by ${\bf c_i}={\bf a_i}+{\bf b_i}$.
At $t_{3N}=0$ and finite
spin-orbit coupling $\lambda_{SO}$, the model Hamiltonian
Eq. (\ref{eqn:GKMham}) is reduced to the Kane-Mele model,
\cite{kane2005a,kane2005b} which is a two-dimensional $\mathbb{Z}_2$
topological insulator.\cite{kane2005a,kane2005b,bernevig2006} 
Like the KM model, the GKM model is invariant under both the time-reversal symmetry and the honeycomb 
space group {\it p6m} symmetry with its point group $D_6$. 

\begin{figure}[!]
\epsfig{file=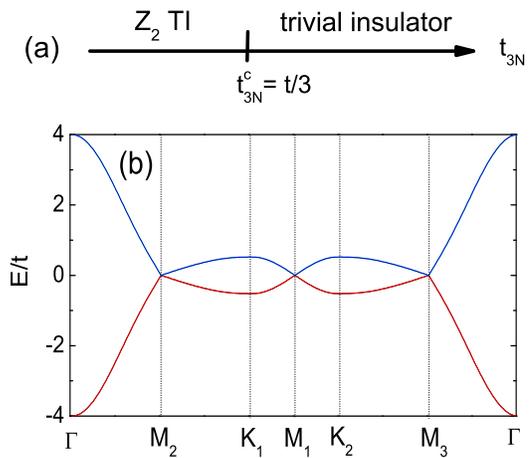,clip=0.1,width=0.8\linewidth,angle=0}
\caption{(Color online) (a) Schematic phase diagram of the GKM model.  (b) The noninteracting band structure
of the GKM model at $t^c_{3N}=\frac{1}{3}t$ (here using
$\lambda_{SO}=0.1t$). The presented momenta are chosen along the
path depicted as blue arrows in Fig. \ref{fig:KMmodel} (b). The gap closes at three TRIM
points: $M_{1,2,3}$, instead of the Dirac point
$K_{1,2}$.}
\label{fig:GKM}
\end{figure}

In the large $t_{3N}$ limit of Eq. (\ref{eqn:GKMham}), the system is a trivial insulator, 
implying the GKM model undergoes a symmetry-preserving topological phase transition as a function 
of $t_{3N}$.\cite{hung2013} The GKM model can be recast as $H_{GKM}=\sum_{\bf k}\Phi^{\dag}_{\bf
k}H^{GKM}_{\bf k}\Phi_{\bf k}$, where $\Phi_{\bf k}=(c_{A\uparrow
{\bf k}},c_{B\uparrow {\bf k}},c_{A\downarrow {\bf
k}},c_{B\downarrow {\bf k}})$ is a 4-component spinor and $H_{\bf
k}$ reads%\cite{ meng2013}
\begin{eqnarray}
H^{GKM}_{\bf k} &=&\left (
\begin{array} {c c c c}
f({\bf k}) & h({\bf k}) &  &  \\
h^*({\bf k}) & -f({\bf k}) &  &  \\
  &  & -f({\bf k}) & h({\bf k}) \\
  &  &  h^*({\bf k}) & f({\bf k})
\end{array}  \right ),\nonumber
\end{eqnarray}
where $h({\bf k})=g({\bf k})-t_{3N}\sum_i e^{i {\bf k} \cdot {\bf
c_i}}$; $g({\bf k})=-t\sum_i e^{i{\bf k}\cdot {\bf a_i}}$, and
$f({\bf k})=2\lambda_{SO}\sum_i\sin{({\bf k}\cdot{\bf b_i})}$; note that ${\bf a_i}$, ${\bf b_i}$ and ${\bf c_i}$ are real-space vectors to describe nearest, second and third-neighbor hoppings. For
most $t_{3N}$ values, the GKM model is gapped. However a simple analysis of the dispersion will show 
a gap closure at $t^c_{3N}=\frac{1}{3}t$ independent of $\lambda_{SO}$, and permits
a change in the topological $\mathbb{Z}_2$ index.  The schematic phase diagram is shown in Fig. \ref{fig:GKM} (a).
For $t_{3N} < t^c_{3N}$, the system is a $\mathbb{Z}_2$ TI, whereas for $t_{3N} > t^c_{3N}$, the system is a trivial insulator. Thus, there exists a topological phase transition at $t_{3N}=\frac{1}{3}t$. 

The non-interacting band structure
of the GKM model is depicted in Fig. \ref{fig:GKM} (b). The chosen momenta are along the high symmetry momentum lines as indicated arrows in Fig. \ref{fig:KMmodel} (b).  Of particular note is the $C_6$ symmetry 
of the dispersion that relates the three high-symmetry $M_{1,2,3}$ points which are also inversion symmetric points. These become \emph{three} Dirac points at the critical topological phase transition; gaps are opened at $K_{1,2}$ due to the spin-orbit coupling $\lambda_{SO}$. One should contrast the topological phase transition with the KM model, which involves \emph{two} Dirac nodes at the $K_1$ and $K_2$ points.  Note that at $t_{3N}=t/3$ and $\lambda=0$, the bands still touch at the $M_{1,2,3}$ points as well as at the $K_{1,2}$ points, yielding 5 Dirac points.

\begin{figure}[]
\epsfig{file=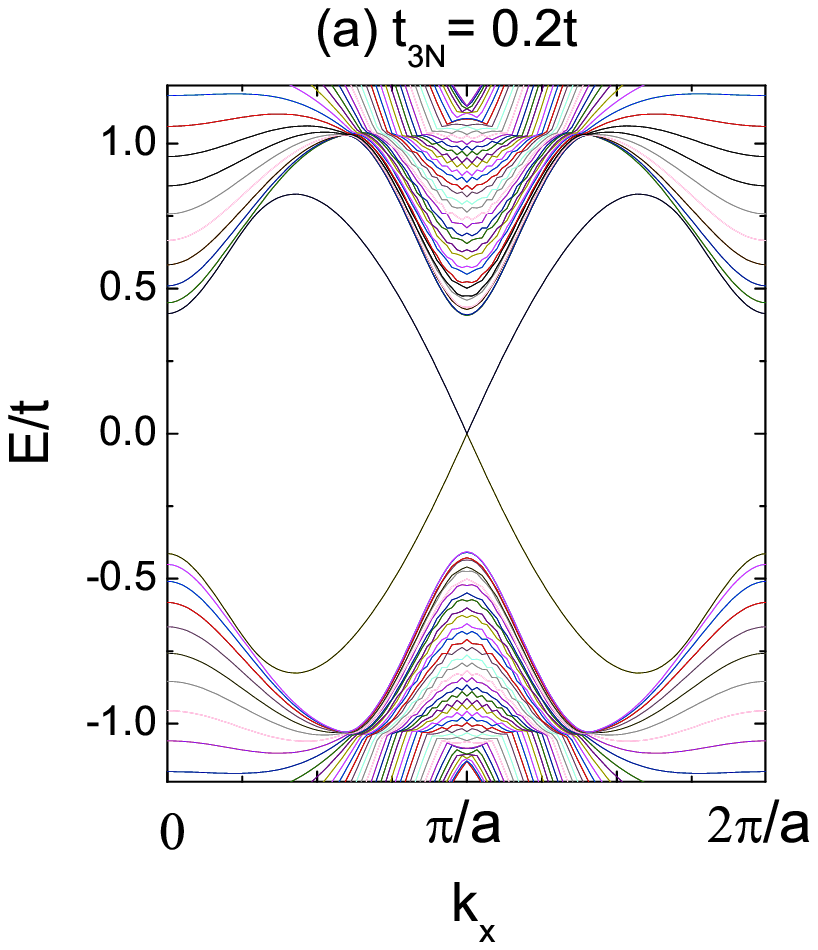,clip=0.1,width=0.5\linewidth,angle=0}
\epsfig{file=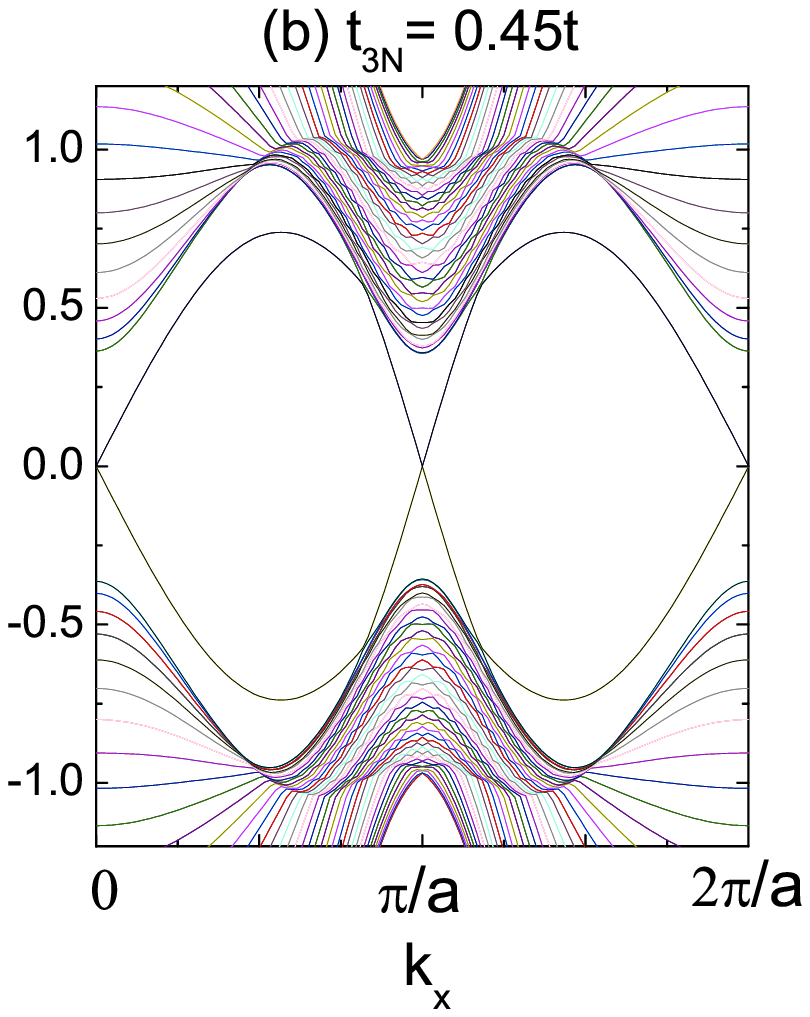,clip=0.1,width=0.465\linewidth,angle=0}
\caption{(Color online) The edge spectra for the noninteracting GKM
model at $\lambda_{SO}=0.2t$ and (a) $t_{3N}=0.2t$, a $\mathbb{Z}_2$
topological insulator and (b) $t_{3N}=0.45t$, a trivial insulator. A
ribbon geometry is used with an armchair edge for periodic boundary
conditions (length, $x$-direction) and zigzag edge for open boundary
conditions (width).} \label{fig:edgestate}
\end{figure}
Displayed in Fig. \ref{fig:edgestate} are results of a band structure computation in a strip geometry of the
GKM model demonstrating the existence of edge states with energies that traverses the bulk energy gap. 
By counting the number of Kramers pairs of edge states per edge,\cite{kane2005a,kane2005b} we note that the GKM model undergoes a topological phase transition as a function of $t_{3N}$, whereby an odd number of Kramers pairs  characteristic of a $\mathbb{Z}_2$ topological insulator phase turns into an even number characteristic of a topologically trivial phase.\cite{kane2005a,kane2005b} Although the two pairs of Kramers helical states in Fig. \ref{fig:edgestate}(b)
shows that the GKM model at $t>t^c_{3N}$ is a $\mathbb{Z}_2$ trivial
insulator,\cite{hung2013} for each spin flavor the spin Chern
number, $C_\sigma$ is even and nonzero. Namely, $|C_{\sigma}|=2 \ne 0$ implying
nontrivial edge states,\cite{ araujo2013} albeit ones not protected by time-reversal symmetry. In addition, since the bulk gap of the GKM model closes at the \emph{three}
time-reversal invariant momenta: $M_1$, $M_2$ and $M_3$ [in Fig. \ref{fig:GKM} (b)], we expect
that the spin Chern number will suffer an \emph{odd} variation\cite{Hatsugai:prb96}
$|\Delta C_{\sigma}|=3$ signaling a topological transition from the $C_{\sigma}=\pm
1$ state to the $C_{\sigma}=\mp 2$ state for spin-up and
spin-down fermions, respectively.  

Note that the appearance of these three Dirac cones, each carrying unit Berry monopole charge, is mandated by the $C_3$ crystal symmetry, since $M_1,M_2,M_3$ transform amongst themselves in a non-trivial irreducible representation of the unbroken $C_3$ symmetry. It must be mentioned that a topological transition with $\Delta C_{\sigma} $ odd may also involve an \emph{even} number of Dirac cones as is the case as in the KM model where a staggered $AB$-site potential competes with spin-orbit coupling. In this instance, however, the inversion symmetry is broken from the pristine graphene band structure allowing the transfer of an odd amount of Chern number. This is because under broken inversion symmetry, the $K_1$ and $K_2$ are not required to contribute equally in the transfer of Chern number. The GKM model, however, differs by always maintaining inversion symmetry and in fact the $M$ points are individually inversion symmetric points. Lastly, from the perspective of the spin Chern numbers, both phases are nontrivial--they exhibit edge robust states so long as $S^z$ symmetry is preserved, and are classified by the spin Chern number.\cite{sheng2006, shengli2005} 

\subsection*{(ii) Dimerized Kane-Mele model}

\begin{figure}[!]
\epsfig{file=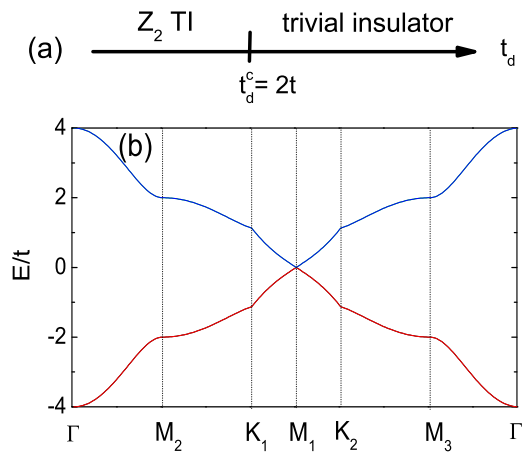,clip=0.1,width=0.8\linewidth,angle=0}
\caption{(Color online) (a) Schematic phase diagram of the DKM model.  (b) The noninteracting band structure
of the DKM model at $t^c_{d}=2t$ (here using
$\lambda_{SO}=0.1t$). The presented momenta are chosen along the
path depicted as blue arrows in Fig. \ref{fig:KMmodel} (b). The gap closes at one TRIM
point: $M_{1}$, instead of the Dirac point $K_{1,2}$.  Compare with Fig.\ref{fig:GKM}.}
\label{fig:DKM}
\end{figure}

The second model we consider, the DKM model is expressed 
by the following Hamiltonian\cite{lang2013}%, meng2013}
\begin{equation}
H_{DKM}  =  - \sum_{\langle
i,j\rangle}\sum_{\sigma}t_{ij}c_{i\sigma}^{\dagger}c_{j\sigma}+i
\lambda_{SO}\sum_{\langle\langle i,j\rangle\rangle}
\sum_{\sigma}\sigma c^{\dagger}_{i\sigma}\nu_{ij} c_{j\sigma},
\label{eqn:dimerham}
\end{equation}
where $t_{ij}=t_d$ when ${\bf r_j}={\bf r_i}+{\bf a_1}$, whereas
$t_{ij}=t$ otherwise. One can recast the Hamiltonian as
$H_{DKM}=\sum_{\bf k}\Phi^{\dag}_{\bf k}H^{DKM}_{\bf k}\Phi_{\bf
k}$, where $H_{\bf k}$ is
\begin{eqnarray}
H^{DKM}_{\bf k} &=&\left (
\begin{array} {c c c c}
f({\bf k}) & h^{\prime}({\bf k}) &  &  \\
h^{\prime}({\bf k})^* & -f({\bf k}) &  &  \\
  &  & -f({\bf k}) & h^{\prime}({\bf k}) \\
  &  &  h^{\prime}({\bf k})^* & f({\bf k})
\end{array}  \right ),\nonumber
\end{eqnarray}
where $h^{\prime}({\bf k})=-t_d e^{i{\bf k}\cdot {\bf
a}_1}-t\sum_{i=2,3}e^{i{\bf k}\cdot {\bf a}_i}$. At $t_d=t$, Eq. (\ref{eqn:dimerham}) is reduced to the KM model. This kinetic Hamiltonian explicitly breaks the $C_3$ subgroup of $D_6$ resulting in the point group $\mathbb{Z}_2^{(m)}\times\mathbb{Z}_2^{(i)}$ which is a mirror reflection perpendicular to $\bf a_1$ and inversion or 180$^{\circ}$ rotation. 
A schematic of its phase diagram and band structure at the critical point are shown in Fig. \ref{fig:DKM}.

\begin{figure}[!h]
\epsfig{file=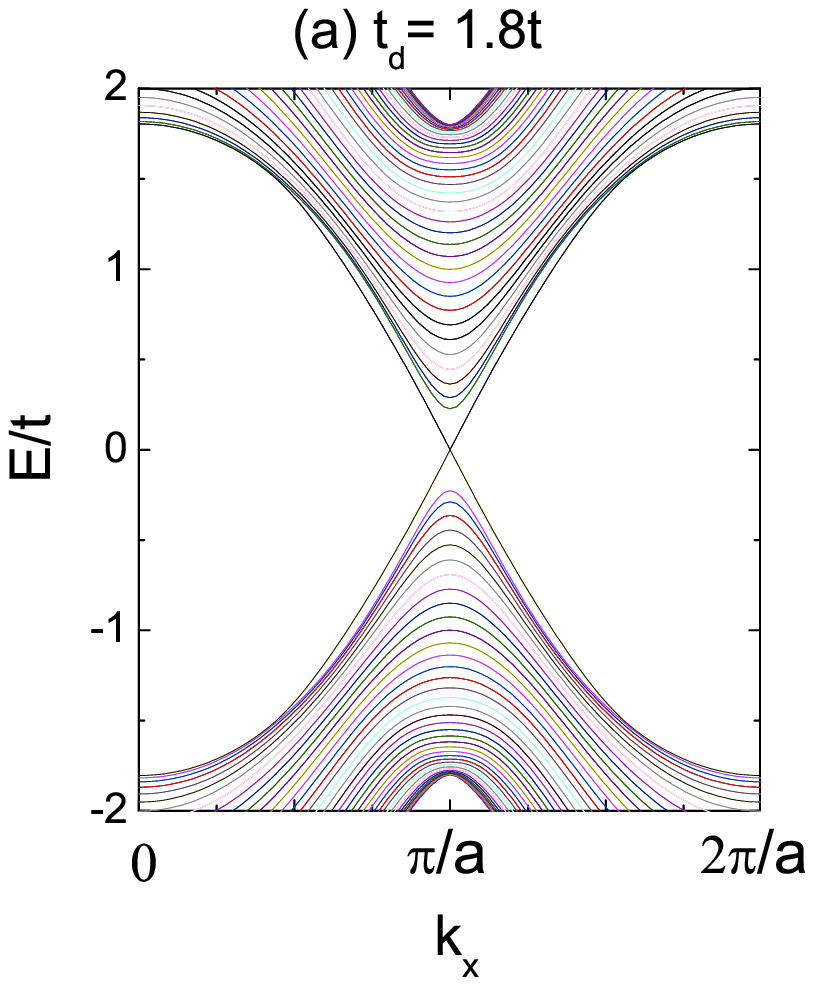,clip=0.1,width=0.5\linewidth,angle=0}
\epsfig{file=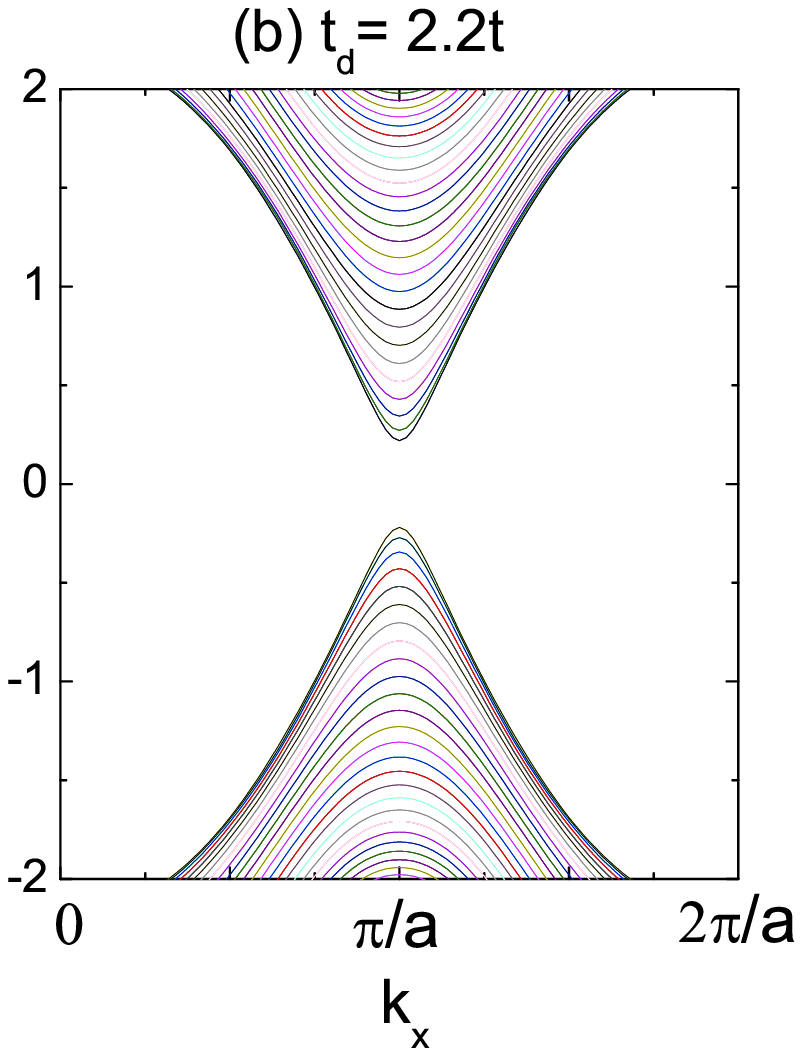,clip=0.1,width=0.46\linewidth,angle=0}
\caption{(Color online) The edge spectra for the noninteracting DKM
model at (a) $t_{d}=1.8t$  and (b) $t_{d}=2.2t$, for a $\mathbb{Z}_2$
topological insulator and a trivial insulator, respectively.
$\lambda_{SO}=0.2t$ is used. The anisotropic hopping $t_d$ is
introduced along the zigzag direction.} \label{fig:edgestate_dimer}
\end{figure}
A topological phase transition will occur by tuning $t_d$ to twice the nearest-neighbor hopping. In this instance the conduction and valence bands touch at a \emph{single} Dirac cone at the $M_1$ point when $t^c_d=2t$ (again independent of the value of $\lambda_{SO}$). This critical point separates a trivial and the topological insulator phase, as shown in Fig.\ref{fig:DKM}.\cite{lang2013} In Fig. \ref{fig:edgestate_dimer} we show the band structure in a strip geometry for the DKM model. The topological phase transition in the DKM model, however, is different from the one in the GKM model as noted by the absence of any helical edge modes on  the trivial insulator side  [shown in Fig. \ref{fig:edgestate_dimer}(b)]. Thus, the trivial insulator phase ($t_d>t^c_d$) has zero spin Chern number and its variation is $|\Delta C_{\sigma}|=1$ during the topological phase transition.

From the symmetry perspective, the transition in the DKM model greatly differs from the GKM model since $C_3$ is completely broken leaving only mirror and inversion $\mathbb{Z}_2^{(m)} \times \mathbb{Z}_2^{(i)} $ symmetries of the original $D_6$ point group. Besides the trivial $\Gamma$ point, the $M_1$ point -- where the \emph{single} critical Dirac cone appears -- is the only inversion symmetric point which also respects the residual mirror symmetry. Thus, the topological phase transition proceeds as a unit change of spin Chern number and hence the topological $\mathbb{Z}_2$ index. In summary, we see that at least in the non-interacting limit, point group symmetry can greatly influence the form of the electronic structure of the critical point straddling a QSH phase and trivial phase.

\subsection*{(iii) $t_L$ Kane-Mele model}

\begin{figure}[!tb]
\epsfig{file=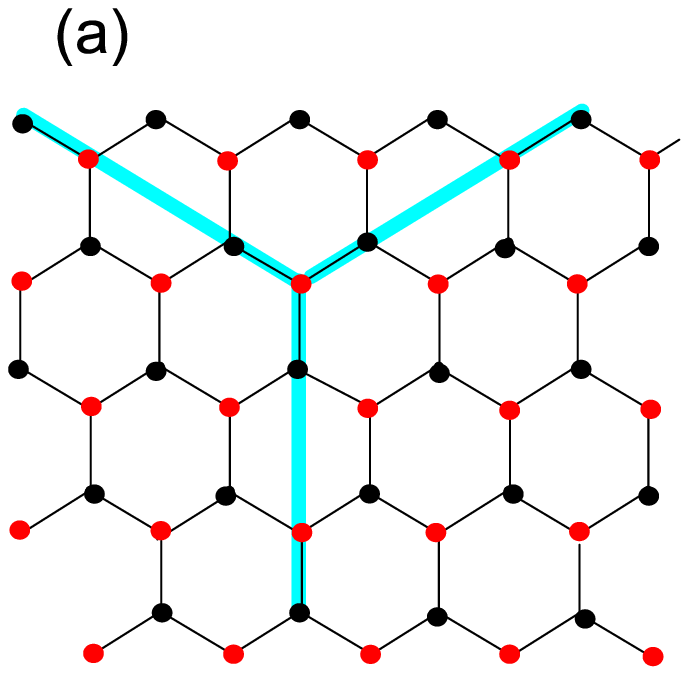,clip=0.1,width=0.45\linewidth,angle=0}
\epsfig{file=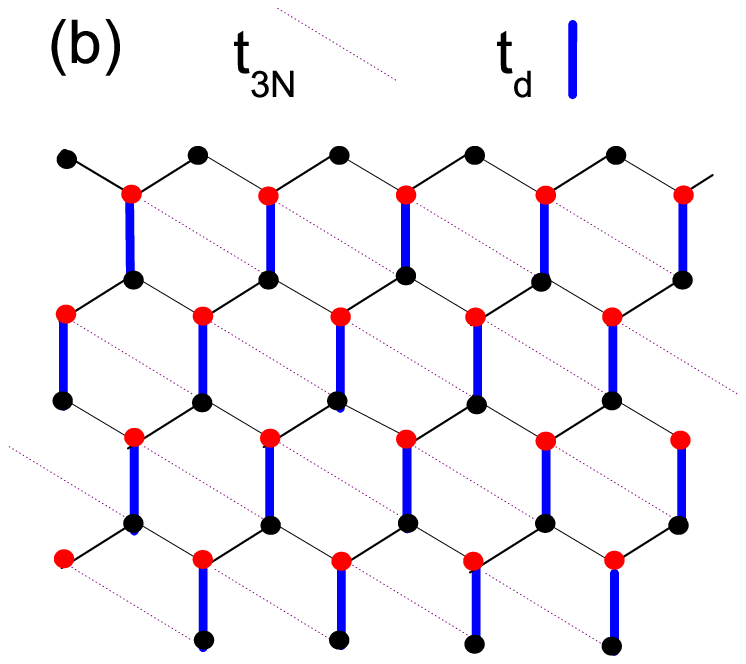,clip=0.1,width=0.5\linewidth,angle=0}
\caption{(Color online) (a) The $t_L$-KM model and (b) The
$t_{3N}$-dimerized KM model.} \label{fig:KMmodelfamily}
\end{figure}

The next model on our list is the $t_L$-KM model which supplements the KM model with a four-lattice-constant-range hopping of strength $t_L$. Similar to the $t_{3N}$ term in the GKM model, the tight-binding parameter, $t_L$, in the $t_L$-KM model preserves the $D_6$ point group symmetry of the honeycomb lattice. The model Hamiltonian reads as
\begin{eqnarray}
 H_{t_L} & = &
- t\sum_{\langle
i,j\rangle}\sum_{\sigma}c_{i\sigma}^{\dagger}c_{j\sigma}+i
\lambda_{SO}\sum_{\langle\langle i,j\rangle \rangle}
\sum_{\sigma}\sigma c^{\dagger}_{i\sigma}\nu_{ij} c_{j\sigma}
\nonumber \\  & &  - t_{L}\sum_{\lbrace
i,j\rbrace}\sum_{\sigma}c_{i\sigma}^{\dagger}c_{j\sigma},
\label{eqn:tLmodel}
\end{eqnarray}
where the first two terms describe the KM model, and in the third
term $\lbrace i,j \rbrace$ denotes the {\em real-valued} hopping with
the distance of $4a$. The lattice structure is shown in Fig.
\ref{fig:KMmodelfamily} (a).
Similar to the GKM model, in the non-interacting limit, there exists
a topological phase transition from the $\mathbb{Z}_2$ topological insulator
to the trivial insulator state. In this instance, the boundary is located at
$t_L=\frac{1}{3}t$ with three $M_1,M_2,M_3$ Dirac cones. For simplicity, we do not discuss the properties of the edge dispersion as they 
are qualitatively similar to the GKM model. 

\subsection*{(iv) $t_{3N}$-Dimerized Kane-Mele model}

The final model we consider is the $t_{3N}$-dimerized KM model which is constructed by
the combination of one third-neighbor hopping (instead of three) \emph{and}
the bond dimerization in the KM model. As shown in Fig. \ref{fig:KMmodelfamily} (b), the solid blue lines denote the
dimerized bonds with $t_d$ strength and the purple dotted lines denote
the diagonal $t_{3N}$ hopping. The Hamiltonian reads as
\begin{eqnarray}
 H_{t_{3N}d} & = &
- \sum_{\langle
i,j\rangle}\sum_{\sigma}t_{ij}c_{i\sigma}^{\dagger}c_{j\sigma}+i
\lambda_{SO}\sum_{\langle\langle i,j\rangle \rangle}
\sum_{\sigma}\sigma c^{\dagger}_{i\sigma}\nu_{ij} c_{j\sigma}
\nonumber \\  & &  - t_{3N}\sum_{\lbrace i,j\rbrace
=\mathbf{c}_3}\sum_{\sigma}c_{i\sigma}^{\dagger}c_{j\sigma},
\label{eqn:dimert3Nmodel}
\end{eqnarray}
where $t_{ij}=t_d$ if $\mathbf{r}_j=\mathbf{r}_i+\mathbf{a}_3$;
otherwise $t_{ij}=t$. The first two terms give the DKM model. The
real-valued diagonal $t_{3N}$ hopping is selected along
$\mathbf{c}_3=\sqrt{3}a\hat{x}-a\hat{y}$. 

The simultaneous presence of the
dimerized bonds and $t_{3N}$ bonds breaks the $\mathbb{Z}_2^{(m)}$ mirror reflection and $C_3$ rotational symmetry of $D_6 \cong (C_3 \rtimes \mathbb{Z}_2^{(m)})\times \mathbb{Z}_2^{(i)}$. Thus the $D_3$ subgroup is completely broken, however the $\mathbb{Z}_2^{(i)}$ inversion symmetry is still respected. There also exists a topological phase transition between the $\mathbb{Z}_2$ topological insulator state and the trivial state, and the non-interacting critical condition can be determined to be $t_{3N}+t_d=2t$ which is again independent of the value of $\lambda_{SO}$. At the topological phase boundary, $t^c_d=2t-t_{3N}$, the bands form a single Dirac cone at $M_3$, which hints  that the spin Chern number has changed by $|\Delta C_{\sigma}|=1$ during the topological phase transition.

%%%%%%%%%%%%%%%%%%%%%%%%%%%%%%%%%%%%%%%%%%%%%%%%%%%%%%%%%%%%%%%%%%%%%%%%%%%%%%%%%%%%%%%

\section{Numerical Evaluation of Topological Indices}
\label{sec:top_ind}

For each generalization or variant of the Kane-Mele model, an on-site Hubbard interaction will be added, and interacting phase diagrams containing the trivial and topological phases are obtained via QMC simulations. But first we review and discuss the quality of numerically computed topological indices of the finite clusters in the non-interacting limit.

The first and most important topological index is the $\mathbb{Z}_2$ invariant of a two-dimensional non-interacting topological insulator.\cite{kane2005b}  When inversion symmetry is present, the noninteracting $\mathbb{Z}_2$ invariant is determined as\cite{ fu2007}
\begin{eqnarray}
(-1)^{\nu}=\prod_{{\mathbf k_i} \in \text{TRIM}} \prod_{m} \xi_{2m}({\bf k_i}),
\label{eqn:noninteractingZ2}
\end{eqnarray}
where $\xi_{2m}(\bf k_i)$ is the parity of $2m$-th occupied Hamiltonian eigenstate at the time reversal
invariant momentum (TRIM); in the KM models, they are $\Gamma$ and $M_{1,2,3}$ as depicted in Fig. \ref{fig:KMmodel} (b). $(2m-1)$-th and $2m$-th states share the same parity and are a Kramers pair, and therefore should only be counted once in the determination of the topological invariant.
The time-reversal invariant topological insulator phase is stable in the weakly interacting limit\cite{ xu2006} and the $\mathbb{Z}_2$ index is also well defined in the case of weak-interactions. It may be obtained conveniently from the zero-frequency single-particle Green's function.\cite{Wang:prx12,Wang:prb12} Specialized to the presence of inversion symmetry, the Fu-Kane\cite{fu2007} expression Eq. (\ref{eqn:noninteractingZ2}) with interactions generalizes to
\begin{eqnarray}
(-1)^{\nu}=\prod_{{\mathbf k_i} \in \text{TRIM}} \tilde{\eta}({\bf k_i}),
\label{eqn:interactingZ2}
\end{eqnarray}
where $\tilde{\eta}({\bf k_i})$ are the parity eigenvalues (one per
Kramer's pair) of the R-zero\cite{ Wang:prx12} \footnote{See the
supplemental material in Ref. [\onlinecite{ hung2013}]} eigenstates
of the zero-frequency Green's functions at TRIM. R- and L-zeros are terms used to refer to eigenfunctions of the zero-frequency single particle Green's function $G_\sigma(i\omega=0,k)$. Eigenfunctions $|v_{nk\sigma}\rangle$ with band index $n$, spin $\sigma$, crystal momentum $k$ and eigenvalue such that 
\begin{align}
\label{eq:G_top}
G_\sigma(i\omega=0,k)|v_{nk\sigma}\rangle =\lambda_{nk\sigma} |v_{nk\sigma}\rangle  
\end{align}
are called R-zeros when  $\lambda_{nk\sigma} >0$, and L-zeros when $\lambda_{nk\sigma}<0$.\cite{Wang:prx12} In the non-interacting limit, R-zeros correspond to occupied states below the Fermi-energy.(Note that Eq.\eqref{eq:G_top} is often expressed in terms of the {\em inverse} Green's function.\cite{ Wang:prx12}  Since our system consists of 2 x 2 matricies for each spin value, we can equivalently express the formula directly in terms of the Green's function.  One need only exercise care in the meaning of L-zeros, R-zeros, and singularities of the Greens functions.)   The singular case $G_\sigma(i\omega,k)\sim 1/\omega$ as $\omega\rightarrow 0$ corresponds to the presence of gapless quasiparticles where a gapped topological insulating phase is not well-defined. The interesting case of $G_\sigma(i\omega=0,k)=0$ \cite{gurarie2011,manmana2012topological,2014arXiv1403.4938Y} or $\lambda_{nk\sigma}=0$ is an indication of the onset of an interaction driven metal-insulator transition in the Brinkman-Rice sense.\cite{brinkman1970application} Either a pole singularity or zero of $G_\sigma(i\omega=0,k)$ may induce a change in the topological index. This expression for the $(-1)^\nu$ index is immensely useful and convenient in determining the topological phase of an interacting time-reversal invariant system, but is however limited to the inversion symmetric situations.

The second topological index that will concern us is the spin Chern number defined by Eq. (\ref{eqn:ChernSpin}) in the QSH context. The Chern numbers $C_\sigma$ of the $S^z$ projected bands are expressed in terms of one particle spectral projectors\cite{PhysRevLett.51.51} as
\begin{align}
C_\sigma = \frac{i}{2\pi}\int d^2k\;
\epsilon^{\mu\nu}\text{Tr}\left[ P_\sigma(k)\;\partial_{\mu}
P_\sigma(k) \;\partial_{\nu}P_\sigma(k) \right],
\label{eqn:c1P}\end{align}  where $P_\sigma(k)=\sum_n
|v_{nk\sigma}\rangle \langle v_{nk\sigma}|$ is the single particle
spectral projector onto R-zero states. Here we have used the Berry curvature in $k-$space interpretation of the spin-Chern number\cite{prodan2009prb} as opposed to the original formulation in terms of twisted boundary conditions.\cite{shengli2005,Sheng:prl03} We will often refer to $C_\sigma$ as the spin Chern number as well, since in the case of $S^z$ conservation--which applies to all cases considered in this work--the spin Chern number is proportional $C_\sigma$, up to a sign determined by convention. More importantly, it is the parity--even or oddness--of $C_\sigma$, not its sign, that determines the time-reversal topological $\mathbb{Z}_2$ index. When $S^z$ is not a good quantum number, the expression Eq. (\ref{eqn:c1P}) for the spin Chern number defined in the thermodynamic limit -- that is without twisted boundary conditions -- may be generalized to the case without $S^z$ conservation rigorously.\cite{prodan2009prb,prodan2010njp}  Even though we will not consider these situations in this work, we would like to point out that it is certainly possible to generalize our numerical methods for the computation of the spin Chern number and hence the $\mathbb{Z}_2$ index for interacting systems where $S^z$ is not conserved. 

The inversion symmetric invariant of Eq. (\ref{eqn:interactingZ2}) and the $S^z$ conserving spin Chern number of Eq. (\ref{eqn:c1P}) exhibit a complementary relationship. The former is only applicable to inversion symmetric Hamiltonians, but does not require the $S^z$ conservation. The latter, however, does not require inversion symmetry but is nevertheless conveniently computed only for $S^z$ conserving Hamiltonians ({\it e.g}. the staggered potential cases\cite{ lai2014}). Moreover, the spin Chern number, which may be any integer value in the thermodynamic limit, carries more information and thus a finer topological classification than the $\mathbb{Z}_2$ index, and can remain quantized even when time-reversal symmetry is broken. However there is an obvious bias towards favoring Eq. (\ref{eqn:interactingZ2}) because by construction it is always integral in finite sized systems. Whilst Eq. (\ref{eqn:c1P}) will in general yield non-integral values in finite-size systems where the Berry curvature over the BZ is no longer smooth. The practicalities of numerically computing the spin Chern number and its sensitivities to finite system size will be the subject of our next discussion. 

For an interacting system, we compute the zero-frequency single-particle Green's function with QMC and then determine its eigenvectors and eigenvalues. The determination of the topological response of a system by the zero-frequency Green's function has been demonstrated in both the non-interacting and interacting limit in
Ref.[\onlinecite{Wang:prx12}]. Both expressions (\ref{eqn:interactingZ2}) and (\ref{eqn:c1P}) sidestep difficulties associated with using twisted boundary conditions,\cite{Niu:prb85} which requires multiple numerically expensive calculations of a non-degenerate ground state. It is also inapplicable when artificial edge degeneracies\cite{Fukui:prb07} are encountered and is usually only practical with exact diagonalization.\cite{Varney:prb11} 

The fact that both of the expressions and their interacting generalizations\cite{Wang:prx12} only rely on the  zero-frequency single-particle Green's functions, is very convenient since more sophisticated numerical simulation methods like QMC and Dynamical Mean-field Theory\cite{ georges1996} (which cannot provide ground state wave functions) can be implemented in determining the topological phases with interactions. The zero-frequency property also implies that numerical analytical continuation does not need to be employed. The computation of Eq. (\ref{eqn:interactingZ2}) for finite-size interacting systems has been previously performed in Refs. [\onlinecite{Go:prl12},\onlinecite{ hung2013}], and is straightforward. There is, however, a requirement that only cluster shapes with BZs containing a TRIM points may be studied with this method. 

\begin{figure}
\epsfig{file=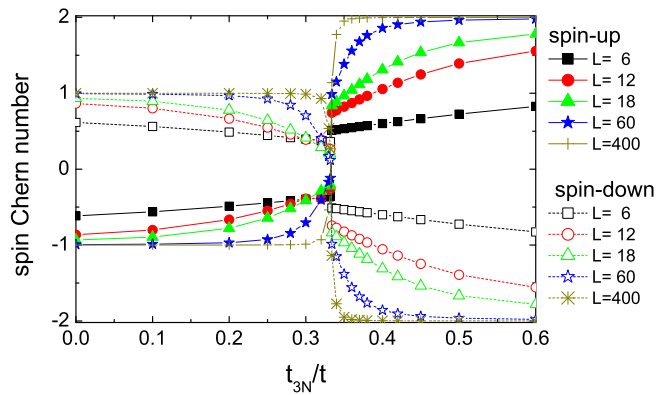,clip=0.1,width=1.0\linewidth,angle=0}
\caption{(Color online) The spin Chern numbers $C_{\sigma}$ vs
$t_{3N}$ for the non-interacting GKM model at $\lambda_{SO}=0.4t$ for
different cluster sizes. For $t_{3N}<\frac{1}{3}t$, the system is a
$\mathbb{Z}_2$ topological insulator with $|C_{\sigma}|=1$. For $t_{3N} >
\frac{1}{3}t$, the system is a trivial insulator but with
$|C_{\sigma}|=2$.} \label{fig:U0chern}
\end{figure}

By contrast computing, Eq. (\ref{eqn:c1P}) for interacting systems is a relatively new enterprise and we describe our numerical method for its computation in finite sizes in Appendix \ref{app:chernnumber}. Our results for the non-interacting GKM model are shown in Fig. \ref{fig:U0chern}.
It is evident that, even in the non-interacting limit, this method of evaluating the spin Chern number
suffers from strong finite-size effects, though it is free of any physical effects associated with twisted boundary conditions. For small sizes, such as $6 \times 6$ and $12 \times 12$, the resulting $C_{\sigma}$ are not well approximated by integer values, but the discontinuous jump at the transition can still be detected by inspection. Only upon increasing the system size do the spin Chern numbers and their discontinuous jumps converge to integers. A finite size scaling analysis is needed to extrapolate to the thermodynamically limit. 

The finite-size effects present in these non-interacting
cases will be important for interpreting the fully interacting
system that we will turn to shortly. However, the precise critical value
of $t^c_{3N}=\frac{1}{3}t$ is clearly seen from the data {\em for all
system sizes} to identify the topological phase transition. In
particular, for the $400 \times 400$ cluster, the variation is
$|\Delta C_{\sigma}|=3$ across the topological phase transition.
As we mentioned earlier, this is consistent with the gap closing at the $M_{1,2,3}$ points.

As was alluded to earlier, the source of the ``non-integerness" of the Chern number is associated with the need to approximate the $k$-space gradients of projector $P_\sigma$ from a finite set of points in the BZ, cf Eq. (\ref{eqn:projector}). This also implies that -- rigorously speaking -- an exact integer value is only ever attainable in the thermodynamic limit. This is an important implication since it means that topological classification as captured by the Chern number and its myriad generalization is an effect that is only rigorously stably protected in the thermodynamic limit. This is intuitively clear since, only in the thermodynamic limit do the energy gaps between smoothly connected Bloch states collapse. The remaining finite energy gaps are the \emph{band gaps} that are the source of the topological protection of a ground state. 

The finite-size computations of Eq. (\ref{eqn:c1P}) shown in Fig. \ref{fig:U0chern} with non-integral results are an honest reflection of the limitation of working with finite-size clusters. We note that an alternative method by Fukui {\it et. al.}\cite{fukui2005} sidesteps this with a construction which always yields an integer result. However this can be misleading since the accuracy of the results requires a critical mesh size, which Fukui {\it et. al}. have estimated. Moreover, the integer values obtained by their method excludes the possibility of using a finite-size scaling analysis to judge the convergence of their results and is a weakness in their method. These considerations also apply to the integral inversion $\mathbb{Z}_2$ invariant $(-1)^\nu$ which should and does fluctuate with cluster size: There is a shift in boundaries based on this invariant with changing cluster size and shape.

The need for large cluster sizes, however, is compensated by the QMC method which provides access to ground state correlators of cluster sizes significantly larger than those manageable by exact diagonalization. Furthermore, it is not necessary to have a manifold of ground states (which also has to be of a sufficient density) as is required with the twisted boundary conditions method. Another important point to note from Fig. \ref{fig:U0chern} is that at a fixed cluster size, the tendency to an integer value improves, the further away the tuning parameter is from the critical point. Furthermore the many-body excitation gap remains open through that portion of parameter space and the single particle Green's function at zero frequency develops no poles or singularities, permitting us to invoke the principle of adiabatic continuity and infer the thermodynamic value of the spin-Chern number of the entire portion of phase space from the finite-size scaling in the large $t_{3N}$ limit and when $t_{3N}=0$. 

\begin{figure}
\epsfig{file=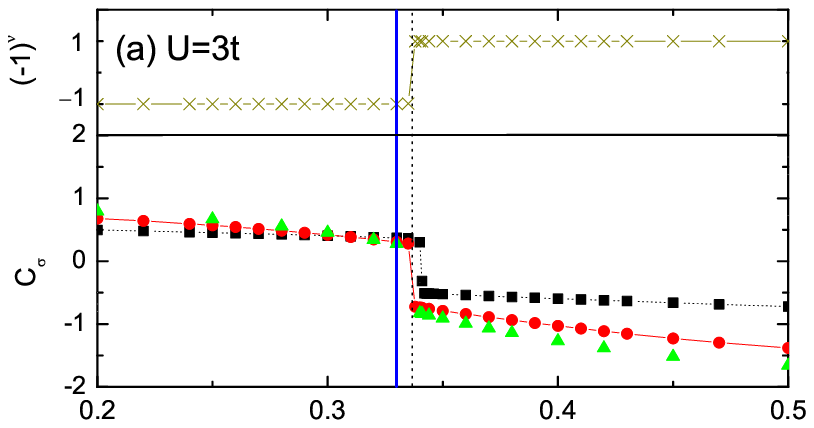,clip=0.1,width=0.95\linewidth,angle=0}
\epsfig{file=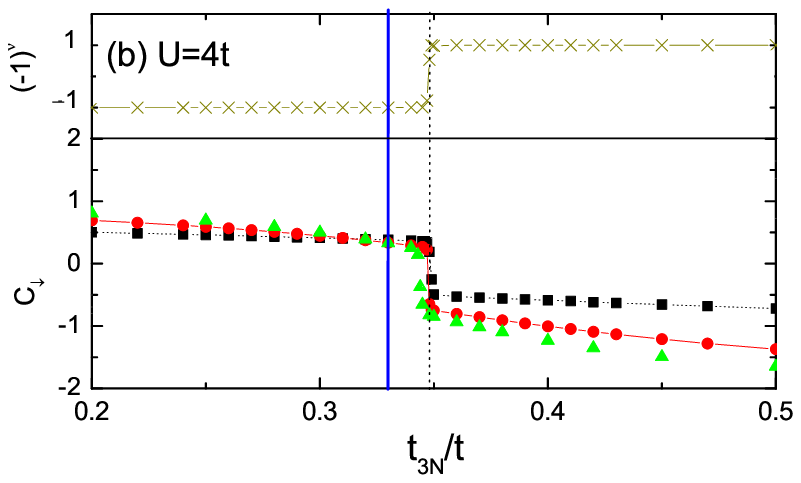,clip=0.1,width=0.96\linewidth,angle=0}
\caption{(Color online) The $\mathbb{Z}_2$ invariant $(-1)^{\nu}$ (upper
panels, for $L=12$ only) and spin Chern number $C_{\sigma}$ (lower
panels) for the GKM-Hubbard model as a function of $t_{3N}/t$ at (a)
$U=3t$ and (b) $U=4t$. The spin-up Chern numbers $C_\sigma$ 
are denoted by solid symbols. The spin-orbit coupling is
$\lambda_{SO}=0.4t$ and the systems sizes are chosen as $6 \times 6$
(black squares), $12 \times 12$ (red circles) and $18 \times 18$ (green triangles).
$t^c_{3N}=\frac{1}{3}t$ (vertical blue line) is the critical point for the non-interacting limit.} \label{fig:GKMspinchern}
\end{figure}

With this information the sudden discontinuity in the numerical Chern number can then be used to pinpoint the critical point. This is the general strategy that we employ in mapping out a phase diagram of both non-interacting and interacting models. In the case of an interacting model phase diagram, we have one more tuning parameter which is the interaction strength itself. The free model can then be trivially classified and when robust excitation gaps persists above the numerical ground state, the principle of adiabatic continuity can be used to reliably map out a phase diagram from sudden jumps in the numerical Chern number.   As a consistency check,  we also compare the spin Chern number with the $\mathbb{Z}_2$ invariant using Eq. (\ref{eqn:interactingZ2}) at various tight-binding parameters and interaction strength, as shown in Fig.\ref{fig:GKMspinchern}.

%%%%%%%%%%%%%%%%%%%%%%%%%%%%%%%%%%%%%%%%%%%%%%%%%%%%%%%%%%%%%%%%%%%%%%%%%%%%%%%%%%%%%%%

\section{Effects of Interactions in Hubbard Model extensions of the Kane-Mele Variants}
\label{sec:int}

We now come to the main part of the paper where we discuss the effects of the Hubbard interaction on interacting topological insulator models. To obtain ground state correlators and capture correlation effects, we use projective quantum Monte Carlo\cite{ sugiyama1986,sorella1989, white1989, assaad2002, assaad2003} to study interacting variants of the KM model, Eq. (\ref{eqn:GKMham})-(\ref{eqn:dimert3Nmodel}) to which an on-site Hubbard term is added, $H \to H +\frac{U}{2} \sum_{i} (n_i-1)^2$ where $U>0$ is the strength of the repulsive on-site Hubbard interaction and $n_i$ is the number operator on site $i$. In our QMC calculations, the number of sites is $N=2\times L^2$, where $L$ takes the values $6$, $ 12$ and $18$.  The largest system sizes are far beyond current capabilities for exact diagonalization studies, rendering our ``unbiased" calculations on interaction effects in topological systems important for going beyond mean-field approaches and the severe finite-size limitations of exact diagonalization studies. The QMC methodology is described in detail in Appendix \ref{app:QMC}.

\subsection*{(i) Generalized Kane-Mele Hubbard model}

We first turn our attention to interaction effects in the GKM-Hubbard model, i.e. $H_{GKM}+U$. Previously in Ref. [\onlinecite{hung2013}], the correlation effects were discovered to result in a shift of the phase boundary that can be accurately computed with QMC simulations: with increasing $U$, $t^c_{3N}$ shifts to larger values (compared to the vertical blue line in Fig.\ref{fig:GKMspinchern}). This behavior was identified by evaluating the $\mathbb{Z}_2$ invariant from exploiting the inversion symmetry of the single-particle Green's function and using Eq. (\ref{eqn:interactingZ2}). The QMC results\cite{ hung2013} showed that at $U=4t$ the topological phase transition boundary moves into the trivial insulator phase by roughly $10\%$. Thus, correlation \emph{stabilizes} the topological phase in the GKM-Hubbard model. 

Here we demonstrate that the topological phase transition can also be clearly identified by computing the spin Chern numbers,\cite{Yoshida:prb13} as shown in the lower panels of Fig. \ref{fig:GKMspinchern}. We chose intermediate interaction strengths $U=3t$ and $U=4t$, which are below the threshold required to induce magnetic ordering or any other symmetry breaking.\cite{hung2013}  For comparison, we also depict the $\mathbb{Z}_2$ invariant vs $t_{3N}$ for the $12\times 12$ cluster. For both $U$ values, Figs.\ref{fig:GKMspinchern} (a) and (b) show marked changes in the spin Chern number at the same locations, as the $\mathbb{Z}_2$ invariant varies for the $12\times 12$ cluster (guided by the dotted lines).

We observe that in the GKM-Hubbard model the QMC sampling still maintains the time-reversal symmetric relation $C_{\uparrow} = -C_{\downarrow}$ within tiny error bars, so long as $t_{3N}$ is far from the phase boundary. When the value of $t_{3N}$ is close to the topological phase transition, one needs to increase the sampling to recover the relation. Similar to the non-interacting limit, the spin Chern numbers in Figs. \ref{fig:GKMspinchern} converge to integers only as $t_{3N}$ is far away from the critical point. 
The spin-up Chern numbers in the $\mathbb{Z}_2$ regime is $C_{\uparrow} \simeq +1$ ($t_{3N}=0.2t$) and turns to $C_{\uparrow} \simeq -2$ after the topological phase transitions ($t_{3N}=0.5t$), indicated in the $12\times 12$ and $18 \times 18$ clusters. The significant variation in $C_{\sigma}$, $|\Delta C_{\sigma}|\simeq 3$, can be used to identify the parameter-driven topological phase transition at finite $U$ in the finite-size clusters. Moreover by adiabatic continuity to the non-interacting limit, we can also confidently identity the two phases between the topological phase transition.

\begin{figure}[!htb]
\epsfig{file=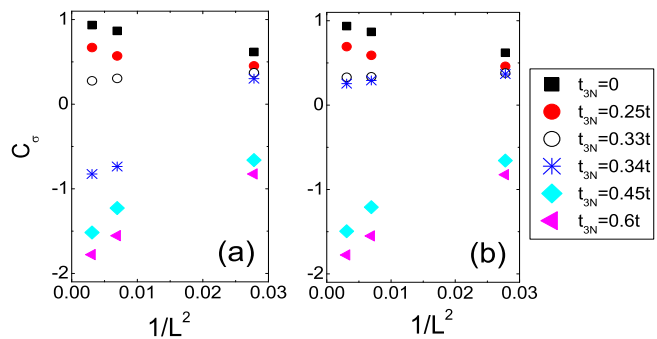,clip=0.1,width=1\linewidth,angle=0}
\caption{(Color online) The tentative finite-size analysis of the
spin-up Chern number $C_{\uparrow}$ of the GKM-Hubbard model at (a)
$U=3t$ and (b) $U=4t$ vs $1/L^2$.} \label{fig:finitesizescaling}
\end{figure}

Next we present the finite-size analysis for the spin Chern
number in the GKM-Hubbard model, shown in Fig.\ref{fig:finitesizescaling}. Since only three different sizes,
$6\times 6$, $12\times 12$ and $18\times 18$ are available, we are
unable to fully capture the scaling behavior. However, the trends are sufficient to infer the value of the thermodynamic spin Chern number. 
On the other hand, the judgement can be also arrived at by the principle of adiabatic continuity to the non-interacting limit of the GKM model. 
We tentatively consider  the Chern number scaling as $1/L^2$ in Fig. \ref{fig:finitesizescaling} 
(or $1/L$, not shown here); showing that as the value of $t_{3N}$ is far away from the
critical point, the spin Chern numbers  extrapolate well to whole
integers, $C_{\uparrow}=1$ or $C_{\uparrow}=-2$ in the thermodynamic
limit.  Thus, these scaling curves are still helpful in
distinguishing the topological states with $C_{\sigma}=1$ and
$C_{\sigma}=-2$. 

Note that near the transition (about $t_{3N}=0.33t$ and
$0.34t$), the scaling analysis is less reliable and one needs bigger
sizes to determine the behavior. However, it is still helpful in
determining the location of the topological phase transition. In
Fig. \ref{fig:finitesizescaling} (a), we can recognize that at
$t_{3N}=0.34t$ the spin Chern number shows a drop with increasing
system size; thus it is a trivial state. By contrast, Fig.
\ref{fig:finitesizescaling} (b) shows that the spin Chern number at
$t_{3N}=0.34$ does not show a clear drop, suggesting that it is still in
the topological insulator regime. Thus, interactions stabilize the
topological phase in the GKM-Hubbard model. Note that for $t_{3N}$
values away from the critical value, the finite size scaling
behavior is much clearer in terms of how the thermodynamic limit is
approached.

Although the values of the spin Chern number suffer from strong finite-size effects, the topological phase transition boundary determined by the topological invariant in the GKM-Hubbard model has weak finite-size dependence. For $U=3t$, on the $6 \times 6$, $12 \times 12$, and $18 \times 18$ clusters, $t^c_{3N}=0.341t$, $0.337t$ and $0.335t$, respectively. For $U=4t$, $t^c_{3N}$ are $0.349t$, $0.347t$ and $0.345t$, respectively, suggesting the spin Chern number is a \emph{reliable} means to detect topological phase transitions in interacting systems.

These interaction effects that cause the critical boundary in phase space to shift must originate from the dynamical quantum fluctuations, since the Hartree-Fock mean-field theory is unable to capture any phase boundary shift (for the $U$ values we consider below the magnetic phase transition).\cite{hung2013,rachel2010} We were not able to develop a perturbative argument for this shift, either.
 %Furthermore, the shift in the phase boundary is perceptibly marginal from $t^c_{3N}=\tfrac{t}{3}$ at $U=0$ limit, even though at $U=4t$ the interaction strength \emph{exceeds} the non-interacting single particle excitation gap of order $O(t)$ . 

\subsection*{(ii) Dimerized Kane-Mele Hubbard model}

We next turn to the DKM-Hubbard model:\cite{ lang2013} $H=H_{DKM}+U$. Recall that at $U=0$, the critical point occurs at $t^c_d=2t$ and is independent of value of $\lambda_{SO}$. Similar to the GKM-Hubbard model, correlation effects induce a shift of the phase boundary, but the critical value of $t_d$ moves \emph{towards} (into) the topological phase.  In other words, correlation \emph{destabilizes} the topological insulator phase--a behavior opposite to the GKM-Hubbard model. With finite interactions at $U=2t$ and $\lambda_{SO}=0.2t$, $t^c_d$ is determined within $1.94$ and $1.96$ by observing the Green's function behavior and the $\mathbb{Z}_2$ topological invariant.\cite{lang2013}

\begin{figure}
\epsfig{file=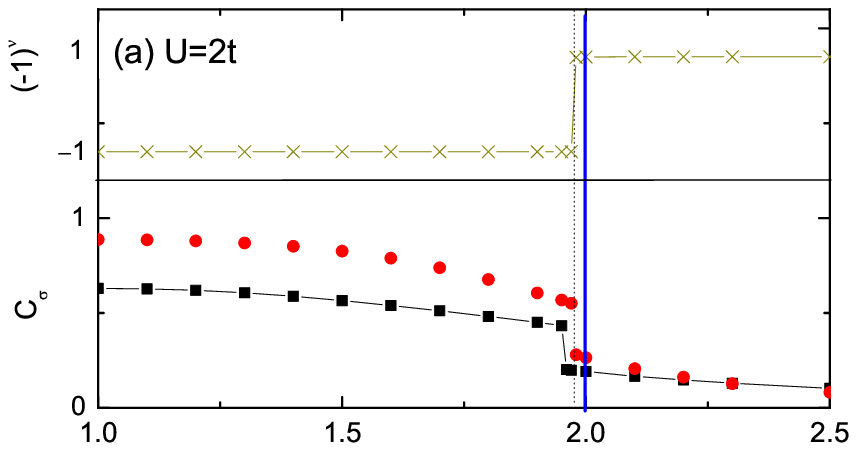,clip=0.1,width=0.95\linewidth,angle=0}
\epsfig{file=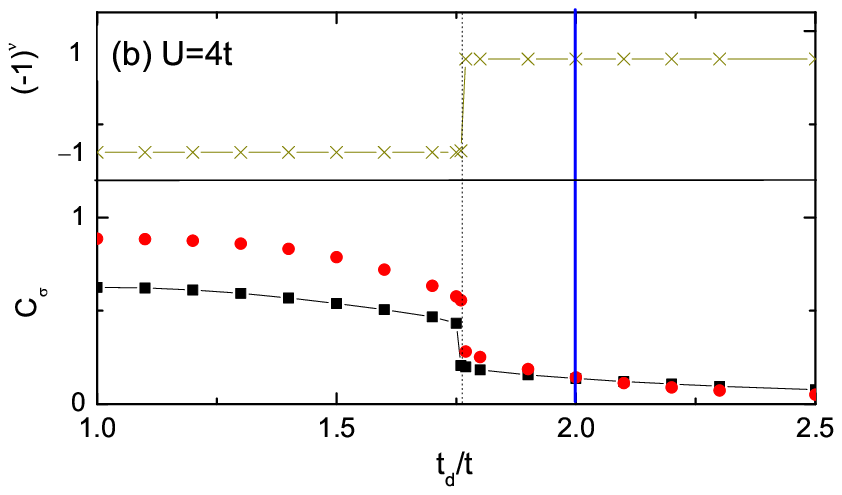,clip=0.1,width=0.95\linewidth,angle=0}
\caption{(Color online) The $\mathbb{Z}_2$ invariant $(-1)^{\nu}$ (upper
panels, for $L=12$ only) and $C_{\sigma}$ (lower panels) for the
DKM-Hubbard model as a function of $t_{d}/t$ at (a) $U=2t$ and (b)
$U=4t$. $\lambda_{SO}=0.2t$ and the systems sizes are chosen as $6
\times 6$ (black squares) and $12 \times 12$ (red circles). At $t_d=t$, the 
system reduces to the standard KM model and $t_d=2t$ (vertical blue line) is the 
critical point for the non-interacting limit. For the sake of clarity,
only the data for $C_{\uparrow}$ is presented here.}
\label{fig:DKMspinchern}
\end{figure}

Here we also employ the QMC combined with the computation of the spin Chern number using Eq. (\ref{eqn:c1P}) and the $\mathbb{Z}_2$ index using Eq. (\ref{eqn:interactingZ2}). Likewise, the values of $U$ we considered were below the magnetic transition. Figs. \ref{fig:DKMspinchern} (a) and (b) show $C_{\uparrow}$ and the $\mathbb{Z}_2$ index vs $t_d$ for $U=2t$ and $U=4t$, respectively.  In the $12\times 12$ cluster (red circles), we can see that the spin Chern number jumps at $t_d=1.97t$ for $U=2t$ and $t_d=1.76t$ for $U=4t$, and, simultaneously, the value of the $\mathbb{Z}_2$ index turns from $(-1)^{\nu}=-1$ to $1$ (guided by the dot lines). More strongly in the $12 \times 12$ cluster, one sees that the topological phase transition occurs between the $|C_{\sigma}|=1$ state to the $|C_{\sigma}|=0$ state and a variation $|\Delta C_{\sigma}|\approx 1$.

Surprisingly, compared to the GKM-Hubbard model, the interaction in the DKM model brings about a more significant shift in the location of the topological phase transition.  In the $L=12$ cluster, $t^c_d$ at $U=2t$ is estimated to be $1.97t-1.98t$, whereas at $U=4t$ it lies within $1.76t-1.77t$. The critical point has shifted by roughly $25\%$. The DKM-Hubbard model also has weaker finite-size effects on the topological phase boundaries. For the $L=6$ cluster, $t^c_d$ s are estimated around $1.95t-1.96t$ and $1.75t-1.76t$ for $U=2t$ and $4t$, respectively. The comparison of the results for the two cluster sizes show similar locations of the topological phase boundaries, thus suggesting weak finite-size effect on the critical points.

\subsection*{(iii) $t_L$- and $t_{3N}$-Dimerized Kane-Mele Hubbard models}

Lastly, we present QMC results for the on-site Hubbard models of the $t_L$-Kane-Mele model Eq. (\ref{eqn:tLmodel}) and the $t_{3N}$-dimerized KM model Eq. (\ref{eqn:dimert3Nmodel}). These two models represent polar opposites with regard to their non-interacting hopping Hamiltonians. The former like the GKM model preserves the full $D_6$ point group, whilst the later breaks it down almost completely to just the inversion subgroup $\mathbb{Z}^{(i)}_2$. Thus, the $t_{3N}$-dimerized KM model is even less symmetric than the DKM model. The motivation for considering these other variants is to demonstrate more examples of interacting TI phases and the role crystal symmetry or lack thereof might play and help contrast the different outcomes of explicitly breaking or preserving the crystal symmetry of the underlying KM model. Given that for the hopping Hamiltonians that we have set out to study, the crystal symmetry already greatly influences the low-energy character of the critical theory -- such as deciding the number of Dirac cones -- between the topological insulator phase and the normal insulator phase, it is reasonable to expect that crystal symmetry will have a significant role to play in shifting phase boundaries. 

\begin{figure}
\epsfig{file=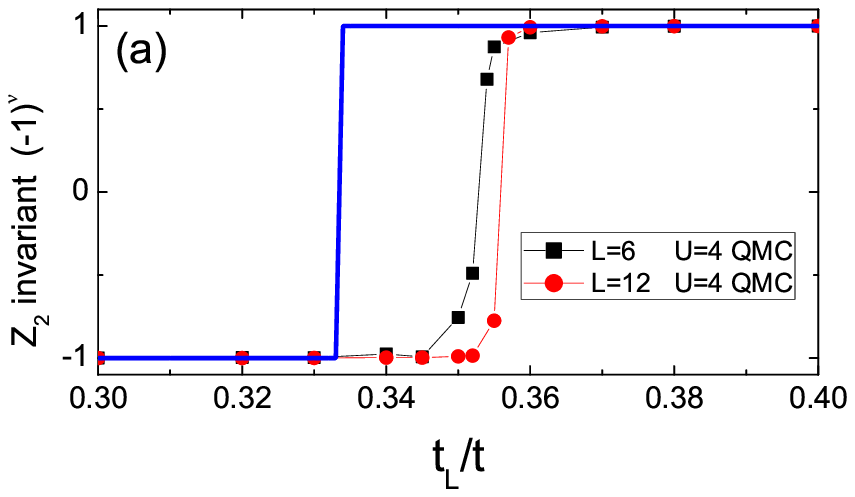,clip=0.1,width=0.85\linewidth,angle=0}
\epsfig{file=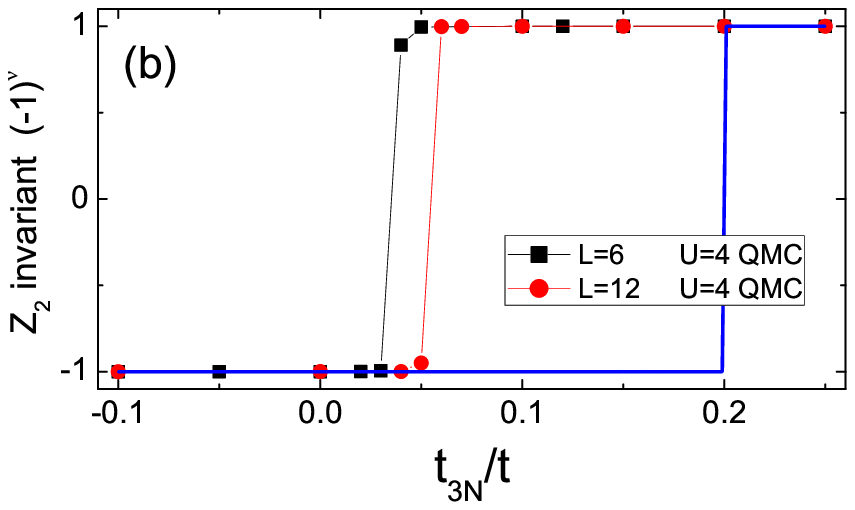,clip=0.1,width=0.82\linewidth,angle=0}
\caption{(Color online) The $\mathbb{Z}_2$ invariants for interacting and
noninteracting cases in  the (a) $t_L$-KM model and (b)
$t_{3N}$-dimerized KM model at $t_d=1.8t$. For reference purposes, the non-interacting $\mathbb{Z}_2$
invariant were computed and presented as the blue lines using $L\times L=1200 \times 1200$ clusters. The
symbols depict the interacting $\mathbb{Z}_2$ invariant by the QMC for
$U=4t$ on $6\times 6$ (black squares) and $12 \times 12$ (red circles). All calculations are
preformed with $\lambda_{SO}=0.4t$. } 
\label{fig:t5_dimert3_Z2}
\end{figure}

The QMC results of the correlation effects on these two models are displayed in Fig. \ref{fig:t5_dimert3_Z2} where we still used $\lambda_{SO}=0.4t$ and $U=4$ to compare with Fig. \ref{fig:GKMspinchern} and Fig. \ref{fig:DKMspinchern}.  For simplicity, we only show the $\mathbb{Z}_2$ invariants as a function of the tight-binding parameters: $t_{L}$ and $t_{3N}$ in Eq. (\ref{eqn:tLmodel}) and Eq. (\ref{eqn:dimert3Nmodel}), respectively. Note that, near the critical point, $(-1)^{\nu}$ shows a poor approximation to an integer value (not $\simeq 1$ or $\simeq -1$), meaning that more QMC samplings are required. However, we still can distinguish the locations of the correlated topological phase boundaries. 
From Fig. \ref{fig:t5_dimert3_Z2} (a), it is clear that with finite interaction, the topological phase transition shifts towards (into) to the trivial insulator regime; the topological phase is enlarged and thus correlation \emph{stabilizes} the topological insulator state in the $t_L$-KM model. For the $6\times 6$ cluster, $t^c_{L}=0.352t-0.354t$, and for the $12\times 12$ cluster, $t^c_{L}=0.352t-0.355t$. Like the GKM model, the $t^c_{L}$ has weak finite-size effect on the phase boundaries. 

In the next panel, Fig. \ref{fig:t5_dimert3_Z2} (b) exhibits the interacting $\mathbb{Z}_2$ invariant against the $t_{3N}$ parameter for the $t_{3N}$-dimerized KM model at $t_d=1.8t$. The non-interacting limit, $t^c_{3N}=0.2t$, is indicated by the blue line. Turning on interaction, the phase boundary moves towards (into) the topological state regime; thus correlation \emph{destabilizes} the topological insulator phase.  We have numerically examined that with finite bond dimerization, the topological critical points are always pushed to the topological insulator regime under correlation.

Our observations of the effects of Hubbard-type interactions on these KM model variants show a systematic pattern: The stability of the topological insulator phase as measured by its occupied volume in the phase diagram is diminished when more of the symmetries of the $D_6$ point group of the lattice are explicitly broken by the Hamiltonian.  A posteriori, we elevate our observations to a speculative conjecture of a principle: in the absence of any spontaneous symmetry breaking, the Hubbard interaction will displace the critical line of the interacting topological quantum phase transition in favor of the normal insulator phase if fewer crystal point group symmetries -- but which must include inversion -- are present in the non-interacting portion of the tight-binding Hamiltonian. The condition regarding spontaneous symmetry breaking excludes competition with magnetic and density-waves phases. This is important to state since at very strong coupling either phase gives way to the N\'eel ordered phase. It is worth reiterating that due to the special form of these Hamiltonians at half-filling, the QMC methodology employed is free of sign-problems and is essentially an exact method for finite clusters up to statistical noise; which can always be systematically improved with greater sampling. In a related study\cite{wuwei2012} of the plaquette KM model, which is not too dissimilar from the ones we have considered, qualitatively consistent results are obtained.

\section{Discussions}
\label{sec:discussion}

We make a few remarks regarding low energy effective theories and present some related speculations.  As was previously mentioned, mean-field calculations at the level of the Hartree-Fock approximation in the KM model\cite{rachel2010} and also our own computations\cite{hung2013} for models Eqs. (\ref{eqn:GKMham})-(\ref{eqn:dimert3Nmodel}) are unable to demonstrate a continuous shift in the topological quantum phase transition boundaries at weak coupling, although a transition to a magnetic state does occur at strong coupling. We rationalize this by noting that the Hubbard $U$ in \emph{two dimensions} for a low energy effective critical field theory of gapless linearly dispersing Dirac fermions is \emph{irrelevant} under scaling. In fact, as is well known\cite{gonzalez1994non} even in the case of long-range Coulomb interactions in graphene, which is a archetype for this variety of field theory, a Renormalization Group (RG) analysis also produces the conclusion that the Dirac nodes are perturbatively stable, albeit with anomalous scaling dimensions due to quantum fluctuations. Thus the phase boundary shifts -- significantly observable only at relatively large $U\sim 3t$ -- that we have observed in our QMC exact computations are effects at  intermediately strong interactions, which is  beyond the weak-coupling low energy-effective theory description. The implications are that the standard field theoretic RG computations at one loop order would be unreliable in capturing the intermediately strong coupling physics of interest. 

Nevertheless, we speculate that an explanation of the dichotomous behavior of the phase boundary shifts must involve the fact that there are a different number of Dirac cones present at the critical topological transition point ({\it three} in the GKM and $t_L$-KM models but {\it one} in the DKM and $t_{3N}$-dimerized KM models) and that this is the main influence of point group symmetry to the low energy physics. It is tempting to relate our observations to a large-$N$ study\cite{herbut2006interactions} of graphene with Hubbard interactions, but we are cautious and reluctant to since $N=1,3$ is a very small value of $N$. In spite of this, Functional Renormalization Group (fRG)\cite{2014arXiv1402.6277J} computations or more recent dimensional regularization $d=3-\epsilon, \epsilon=1$ studies\cite{herbut2009relativistic,assaad2013pinning} applied to the Hubbard graphene system have been encouraging in describing physics near strong coupling. We will leave these very interesting lines of investigations for future work, as these computations are by no means trivial undertakings. 

\section{Summary and Conclusions}
\label{sec:summary}

In this work we have analyzed variants of the archetypical model of a time-reversal symmetric topological insulator, the Kane-Mele model. These generalized models Eqs. (\ref{eqn:GKMham})-(\ref{eqn:dimert3Nmodel}) exhibit various space group symmetries of the honeycomb lattice on which they are formulated. In the non-interacting limit, all of the models exhibit a topological phase transitions between a $\mathbb{Z}_2$ topological insulator phase and a normal insulating phase. By means of the unbiased QMC method, we further study the interacting variants of these model systems by including on-site Hubbard interactions. The projective determinant QMC method that we employ is free of sign problems (at half-filling which is the only filling considered) and is essentially exact up to statistical sampling noise. The regime that interests us most is the intermediately strong $U$ regime before magnetic order sets in. We demonstrate that the topological phase of our numerically exact interacting ground states can be ascertained by computing either the $\mathbb{Z}_2$ invariant or the spin Chern number $C_\sigma$ via the zero-frequency single-particle Green's function. Thus our work is a numerical implementation of the theoretical proposal of Refs. [\onlinecite{Wang:prx12,Wang:prb12}] for finite-size clusters using reliably accurate QMC. The spin Chern number had not been previously computed with QMC, and we argue on technical grounds that it is complementary to the inversion symmetry based expression for the $\mathbb{Z}_2$ invariant. 

Accompanied with finite-size scaling analyses and adiabatic continuity to non-interacting limits, -- which we numerically observe -- we argue that the spin Chern number is a robust classification method of interacting TI's. Moreover the spin Chern number may be utilized in circumstances where inversion and even time-reversal symmetry are absent, and may be generalized to the case where $S^z$ conservation is absent.\cite{prodan2009prb} Our numerically exact QMC results suggest that quantum fluctuations from intermediately strong interactions can act to either stabilize or destabilize the topological phase, depending on whether the hopping terms preserve or break lattice symmetries (when $U$ is less than the value which induces magnetism). Although admittedly bold and speculative, we conjecture a principle that in the situations where spontaneous symmetry breaking phases are excluded, on-site Hubbard interactions will destabilize the TI phase in honeycomb models when the lattice point group $D_6$ is explicitly broken down to a subgroup containing $\mathbb{Z}_2^{(i)}$ inversion by the tight-binding Hamiltonian, and stabilized when the full $D_6$ symmetry is preserved. We speculate that the mechanism by which this acts is through influencing the form of low energy theory at the quantum critical point -- which needs to be handled beyond perturbatively weak coupling -- and suggest further avenues of investigation. We hope our work will help stimulate further studies in this direction and provide a baseline for the general expectations for unbiased calculations of correlation effects on topological phase transitions. 

We are grateful for discussions with Q. Niu, D. N. Sheng, K. Sun, C.
Varney, Z. Wang, S. Yang, W-C Lee. This work was supported by ARO Grant No.
W911NF-09-1-0527, NSF Grant No. DMR-0955778, and by grant
W911NF-12-1-0573 from the Army Research Office with funding from the
DARPA OLE Program.   Simulations were run on the Texas Advanced
Computing Center (TACC) at the University of Texas at
Austin,http://www.tacc.utexas.edu, and the Brutus cluster at ETH
Zurich. VC gratefully acknowledges the support from the DOE grant DE-FG02-07ER46453. 

%\bibliography{/Users/wanglei/Documents/Papers/papers}

\bibliography{KM,pyro111}

\appendix

\section{Quantum Monte Carlo}\label{app:QMC}

The projective quantum Monte Carlo  method (QMC) is given by
projecting an arbitrary trivial wave function $|\psi_T\rangle$
(requiring $\langle \psi_T |\psi_0\rangle \ne 0$) onto the ground
state wave function
$|\psi_0\rangle$.\cite{sugiyama1986,sorella1989,white1989,assaad2002,assaad2003,meng2010,hung2011}
The expectation value of an observable $A$ is obtained by
\begin{eqnarray}
\langle A\rangle =\lim_{\Theta \to \infty}\frac{\langle \psi_T |e^{-
\frac{\Theta }{2}H} A e^{- \frac{\Theta
}{2}H}|\psi_T\rangle}{\langle \psi_T |e^{-\Theta H}| \psi_T\rangle},
\label{eqn:expection}
\end{eqnarray}
where $\Theta$ is the projective parameter. To carry out the
procedures numerically, we need to discretize the projection
operator $e^{-\Theta H}$ into tiny time propagators $e^{-\Delta \tau
H}$ with $\Theta=\Delta \tau M$: $e^{-\Theta H}=(e^{-\Delta \tau
H})^M$ where $M$ is the number of time slices and $\Delta \tau$ is
chosen as a small number. The first-order Suzuki-Trotter
decomposition can further decompose $e^{-\Delta \tau H}$ as
\begin{eqnarray}
e^{-\Delta \tau H} \simeq e^{-\Delta \tau H_0}e^{-\Delta \tau H_U},
\label{eqn:suzuki}
\end{eqnarray}
where $H_0$ is the tight-binding Hamiltonian, which could be equation
(\ref{eqn:GKMham}) and equation (\ref{eqn:dimerham}), for the GKM model
and the DKM model, respectively; $H_U=\frac{U}{2}\sum_i(n_i-1)^2$ is
the repulsive Hubbard on-site interaction;
$n_i=\sum_{\sigma}c^{\dag}_{i,\sigma}c_{i,\sigma}$. To represent
$e^{-\Delta \tau H_U}$ in terms of the single-particle basis, we need to
implement the $SU(2)$-invariant Hubbard-Stratonovich
transformation\cite{ meng2010}
\begin{eqnarray}
e^{-\Delta \tau
\frac{U}{2}(n_i-1)^2}=\frac{1}{4}\sum_{l=\pm1,\pm2}\gamma(l)e^{i\sqrt{\Delta
\tau \frac{U}{2}}\eta(l)(n_i-1)}+O(\Delta
\tau^4),\label{eqn:HStransformation}
\end{eqnarray}
where $\gamma(\pm1)=1+\sqrt{6}/3$, $\gamma(\pm2)=1-\sqrt{6}/3$;
$\eta(\pm1)=\pm\sqrt{2(3-\sqrt{6})}$ and
$\eta(\pm2)=\pm\sqrt{2(3+\sqrt{6})}$ are 4-component auxiliary
fields.  In the current literature, $\Delta \tau t=0.05$ and $\Theta
t=40$ are used through the content.

Implementing equation (\ref{eqn:suzuki}) and
(\ref{eqn:HStransformation}), $H$ turns out to be $\tau$-dependent
since $H_U$ is associated with the auxiliary field configuration
$\eta(l_{i,\tau})$; then $e^{-\Theta H }=\prod^M_{\tau=1}e^{-\Delta
\tau H_{\tau}}$. The denominator of equation (\ref{eqn:expection}) (named the projector partition function),\cite{sorella1989} is evaluated
as follows \cite{assaad2002,hohenadler2012,zheng2011}
\begin{widetext}
\begin{eqnarray}
\langle \psi_T | e^{-\Theta H} |\psi_T\rangle &=& \langle \psi_T |
\prod^M_{\tau=1} e^{-\Delta \tau H_{\tau}} |\psi_T\rangle \cong
\langle \psi_T | \prod^M_{\tau=1} e^{-\Delta \tau H_0}e^{-\Delta
\tau
H_{U,\tau}} |\psi_T\rangle\nonumber\\
&=&(\frac{1}{4})^{MN}\sum_{\lbrace l_{i,\tau} \rbrace} \Bigg\{ \Big(
\prod_{i,\tau} \gamma(l_{i,\tau})\Big) \prod_{\sigma} \textrm{Tr}
\Big(\prod^M_{\tau=1} e^{-\Delta \tau
\sum_{i,j}c^{\dag}_{i,\sigma}[{\bf
H^{\sigma}_{0}}]_{ij}c_{j,\sigma}}e^{i\sqrt{\Delta \tau
\frac{U}{2}}\eta(l_{i,\tau})(n_{i,\sigma}-\frac{1}{2})}
\Big)\Bigg\}\nonumber\\
&=&(\frac{1}{4})^{MN}\sum_{\lbrace l_{i,\tau} \rbrace} \Bigg\{ \Big(
\prod_{i,\tau} \gamma(l_{i,\tau}) \Big) \det
\Big(O_{\uparrow}[\eta(l_{i,\tau})] \Big)\det
\Big(O_{\downarrow}[\eta(l_{i,\tau})] \Big) \Bigg\},
\end{eqnarray}
\label{eqn:paritionfun}
\end{widetext}
where $\sum_{l_{i,\tau}}$ runs over possible auxiliary
configurations $\eta(l_{i,\tau})$, where $i=1 \sim N$ are site
indices and $\tau=1 \sim M$ are imaginary time indices; ${\bf
H^{\sigma}_0}$ is the matrix kernel of $H_0$ with spin $\sigma$.
Each time propagator $e^{-\Delta \tau H_0}e^{-\Delta \tau H_U}$ is a
$N \times N$ matrix and $\textrm{Tr}(\prod_{\tau}
e^{\cdots})=\textrm{det}(O_{\sigma})$ represents the trace over
fermion degrees of freedom.

Given such a $N$-site and $M$-time slice system, the summation in
the above equation has a degree of $4^{NM}$, and generally, it is
impossible to consider all configurations. The auxiliary field
configuration, $\lbrace \cdots \eta(l_{i,\tau}) \cdots \rbrace$,
however, can be determined by Monte Carlo importance
samplings.\cite{assaad2002,assaad2003,hung2011} For simplicity, we
used the Metropolis algorithm in this paper.\cite{metropolis1953}
The physical meaning of $\Big( \prod_{i,\tau} \gamma(l_{i,\tau})
\Big)\prod_{\sigma=\uparrow,\downarrow}\textrm{det}\Big(O_{\sigma}[\eta(l_{i,\tau})]\Big)$
is the probability weight at the given auxiliary field configuration
$\lbrace \eta(l_{i,\tau})\rbrace$.\cite{hirsch1985} When this term
is proven positive-definitive, QMC simulations are free-sign and the
results are numerically exact. This is always true in the
half-filling Kane-Mele-type model without Rashba spin-orbital
coupling.\cite{zheng2011,hohenadler2012,hung2013,lang2013}

To obtain zero-frequency Green's function, we first evaluate the
time-displaced Green's function $G({\bf r},\tau)$. The unequal-time
Green's function is defined as\cite{assaad1996}
\begin{eqnarray}
G_{\sigma}(\tau,r_i,r_j)&=&\langle \psi_0 |
c_{\sigma}(\tau,r_i)c^{\dag}_{\sigma}(r_j)|\psi_0\rangle \nonumber \\
&=&\langle \psi_0 |e^{\tau H} c_{\sigma}(r_i)e^{-\tau
H}c^{\dag}_{\sigma}(r_j)|\psi_0\rangle. \label{eqn:unequaltimegf}
\end{eqnarray}
Then we perform the Fourier transform from real space to momentum space
${\bf r} \to {\bf k}$, and imaginary time to the Matsubara frequency
$\tau \to i\omega$,
\begin{eqnarray}
G_{\sigma}(i\omega,{\bf k})&=&\frac{1}{\beta}\int^{\beta}_0 d \tau
e^{i\omega \tau}\frac{1}{N}\sum_{r_i,r_j} e^{i{\bf k}\cdot
(r_i-r_j)}G_{\sigma}(\tau,r_i,r_j).\nonumber
\end{eqnarray}
The zero-frequency is given setting $i\omega=0$.  To calculate the
spin Chern number $C_{\sigma}$, however, we need the single-particle
Green's functions for all momentum points and implement equation
(\ref{eqn:c1P}). This procedure is slightly different from the
approach to evaluate the $\mathbb{Z}_2$ index, for which only time-reversal
invariant momentum points are required.\cite{Wang:prx12} For
sign-free QMC simulations, one can accurately calculate the
zero-frequency Green's functions in system sizes which are larger
than the small clusters in an exact diagonalization and then evaluate
the spin Chern numbers using the projection operators equation
(\ref{eqn:c1P}). This approach is useful to identify different
topological phases in the interacting level without using twisted
boundary conditions.

Note that for more generic cases, the Green's functions are a $4\times 4$ matrix\cite{ sheng2006}, i.e.,
\begin{eqnarray}
\mathbf{G} &=&\left (
\begin{array} {c c}
G_{\uparrow \uparrow} & G_{\uparrow \downarrow} \\
G_{\downarrow \uparrow} & G_{\downarrow \downarrow}
\end{array}  \right ),
\end{eqnarray}
where $G_{\uparrow \uparrow}=G_{\uparrow}$ ($G_{\downarrow
\downarrow}=G_{\downarrow}$) as defined in equation
(\ref{eqn:unequaltimegf}), and $G_{\uparrow \downarrow}=\langle
c_{\uparrow}(\tau) c^{\dag}_{\downarrow}\rangle$. Without Rashba
spin-orbital coupling, however, the $G_{\sigma \sigma^{\prime}}=0$
for $\sigma \ne \sigma^{\prime}$. Therefore, for the simplified
Kane-Mele-type model, the Green's functions reduce to $2 \times 2$
matrix for each spin. In the main text, we implement the QMC and
projection operator procedures equation (\ref{eqn:c1P}) on the GKM and
DKM model to study the parameter-induced topological phase
transition.

 In the noninteracting KM models, due to the inversion symmetry,
the $2\times 2$ Green's functions at the time-reversal invariant
momentum points (TRIM) can be simply expressed as
\begin{eqnarray}
G_{\uparrow \uparrow}(i\omega=0,{\mathbf k}_{i})=\alpha_{\mathbf
k_i} \sigma^x, \ {\mathbf k}_i \in \textrm{TRIM},\label{eqn:gamma}
\end{eqnarray}
where some coefficients multiply the $\sigma^x$ Pauli matrix,
and in equation (\ref{eqn:interactingZ2}),
$\tilde{\eta}({\mathbf{k}_i})=\pm 1$ is well-defined.  In the cases
of finite $U$,  $ \tilde{\eta}({\mathbf k}_i) =\pm1$ and the
relation equation (\ref{eqn:gamma}) are not guaranteed in a single
measurement in the QMC simulations, however. Instead, they should be
obtained by sufficiently large number of QMC simulations.

\begin{figure}[!htb]
\epsfig{file=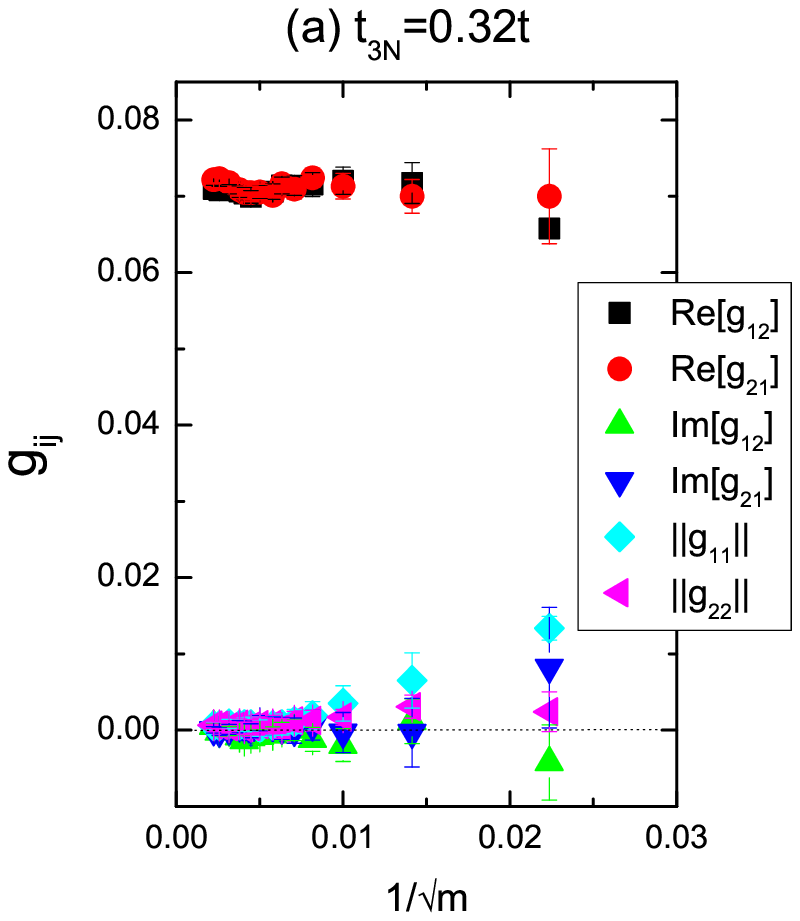,clip=0.1,width=0.46\linewidth,angle=0}
\epsfig{file=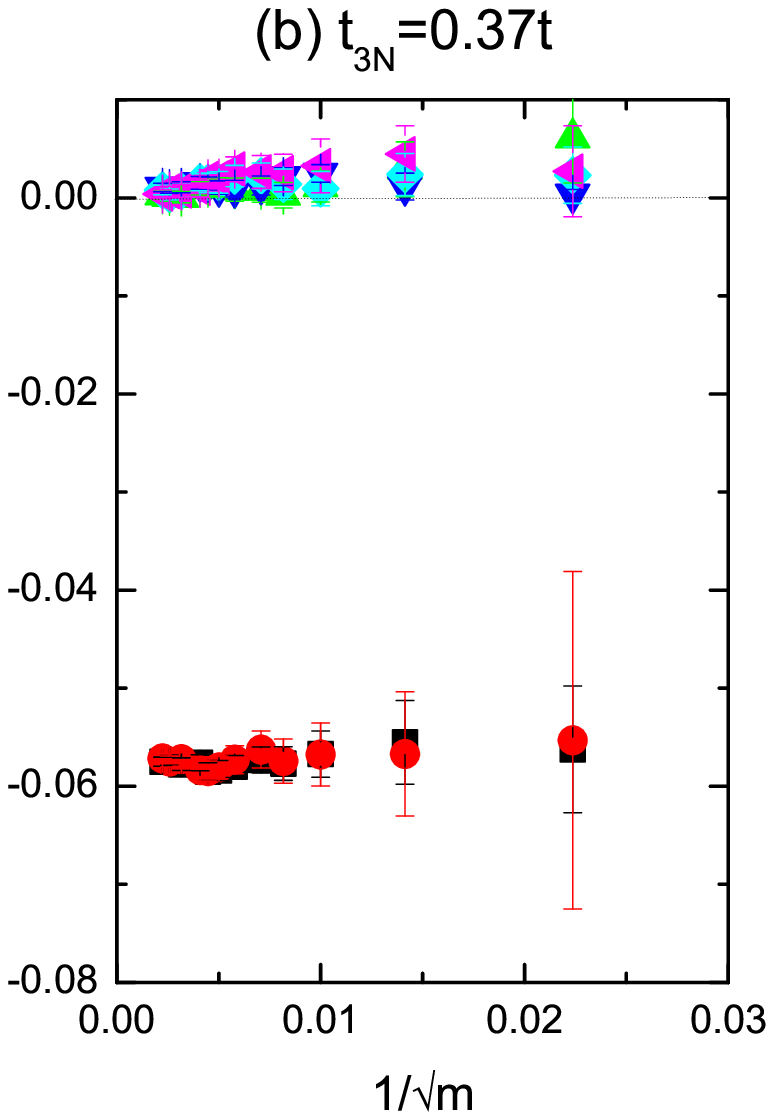,clip=0.1,width=0.37\linewidth,angle=0}
\caption{(Color online) The matrix elements of the zero-frequency
Green's functions $G(0,{\mathbf M_1})$  vs the number of samplings
$m$ at (a) $t_{3N}=0.32t$ and (b) $t_{3N}=0.37t$.
$\lambda_{SO}=0.4t$ and $U=4t$. $\textrm{Re}[g_{ij}]$ and
$\textrm{Im}[g_{ij}]$ denote the real part and imaginary part of
$[G(0,{\mathbf M_1})]_{ij}$, respectively; $||g_{ii}||$ denotes the
diagonal component of $G(0,{\mathbf M_1})$ in magnitudes.}
\label{fig:benchmark}
\end{figure}

To interpret this, we present two  benchmark results for the matrix
elements of the zero-frequency Green's functions at ${\mathbf
k_i=M_1}$, $g_{ij}=[G(i\omega=0,{\mathbf M_1})]_{ij}$, vs the number
of measurements ($m$) in Figs. \ref{fig:benchmark}.
$\lambda_{SO}=0.4t$ and $U=4t$ are used. The test system size is
$2\times 6^2$. To recover equation (\ref{eqn:gamma}), one should expect
that $\textrm{Re}[g_{12}] \simeq \textrm{Re}[g_{21}]$, and
$\textrm{Im}[g_{12}]=\textrm{Im}[g_{21}]=||g_{11(22)}||\simeq 0$. It
has been demonstrated that the values of
$\textrm{Re}[g_{12}]=\alpha_{\mathbf k_i}$ can be used to identify
the topological property\cite{ hung2013}.

Fig. \ref{fig:benchmark} (a) shows $t_{3N}=0.32t$ in the $\mathbb{Z}_2$
topological insulator phase and (b) for $t_{3N}=0.37t$ in the
trivial insulator.  At small $m$, the real parts of $g_{12}$ and
$g_{21}$ are not equal; furthermore, $g_{12}$ and $g_{21}$ have
imaginary parts, and both of $g_{11}$ and $g_{22}$ are finite.
However, one can see that, upon sampling sufficient times,
$\textrm{Re}[g_{12}] \simeq \textrm{Re}[g_{21}]$, and meanwhile
$\textrm{Im}[g_{12}]$, $\textrm{Im}[g_{21}]$, $||g_{11(22)}||$ go to
zero. Thus, in the $m \to \infty$ limit, equation (\ref{eqn:gamma}) is
recovered, and then the value of $\langle
\tilde{\eta}({\mathbf{k}_i}) \rangle$ over QMC simulations
monotonically approaches to $\pm 1$.

For other ${\mathbf k}_i$ and interacting case, equation
(\ref{eqn:gamma}) does not hold. However, for the non-interacting case we found that the value of
resulting spin Chern number $C_{\sigma}$ is not sensitive to the number of
samplings provided they are large enough in number. Throughout our paper, we choose the number of
measurements large enough (mostly over several thousands) to
determine the $2\times 2$ single-particle Green's function, and then
calculate the spin Chern numbers.

\section{Projection operator expression of the Spin-Chern number}\label{app:chernnumber}

In this section we provide a description of the projection operator
expression used to evaluate the spin Chern number for finite lattices and its practical numerical implementation.
The expression for the Chern numbers using the projection
operators onto the occupied bands is,\cite{PhysRevLett.51.51}
\begin{eqnarray}
C_{\sigma}&=&\frac{1}{2\pi i}\int_{\text{B.Z.}}\text{Tr}\, \left(
P_{\sigma} dP_{\sigma} \wedge dP_{\sigma} \right) \nonumber \\  &=&
\frac{1}{2\pi i}\int_{\text{B.Z.}} \text{Tr}\Big\lbrace
P_{\sigma}({\bf k}) \Big[
\partial_{k_x} P_{\sigma}({\bf k}) \partial_{k_y}P_{\sigma}({\bf k}) \nonumber \\  & &
- \partial_{k_y} P_{\sigma}({\bf k}) \partial_{k_x}P_{\sigma}({\bf
k}) \Big] \Big\rbrace dk_x dk_y,\label{eqn:chernnumb}
\end{eqnarray}
where $P_{\sigma}({\bf k})$ is the spectral projector operator
constructed using the Bloch eigenvectors (eigenspace) at ${\bf k}$
with energies below the Fermi energy $\epsilon_F$, i.e., $E_n({\bf
k})<\epsilon_F$ and for spin sector-$\sigma$. A merit of this formulation of the Chern number is the manifest independence of the U(1) phases of the Bloch states. The Bloch eigenstates themselves
are obtained from diagonalizing the interacting zero-frequency Green's functions
\begin{eqnarray}
G_{\sigma}({\bf k},0)|\mu_i \rangle=\mu_i |\mu_i \rangle,
\end{eqnarray}
and then
\begin{eqnarray}
P_{\sigma}({\bf k})=\sum_{\mu_i>0}|\mu_i\rangle\langle\mu_i|,
\end{eqnarray}
where choosing $\mu_i>0$ corresponds to selecting occupied bands
$E_n<\epsilon_F$, i.e. R-zero of the $G_{\sigma}({\bf k},0)$. The
projection operator formula above is manifestly $U(1)$ gauge
invariant. The integral is over the Brillouin zone (BZ), and, in
practical numerical application, the region of integration over the
BZ does not need to be a Wigner-Seitz unit cell in reciprocal
lattice space, as long as the entire reciprocal lattice unit cell is
covered.

In a finite-size system,  the set of $\bf k$-vectors is discretized,
so we will need to replace the integral with the summation over
finite momentum points. For convenience we can map the momentum
points as a $N=L_x \times L_y$ square grid  of spacing $h$ and label
each $\mathbf{k}$ with discrete coordinate indices $\lbrace m,n
\rbrace$. Then we can approximate the partial derivatives
$\partial_{k_x} P_{\sigma}({\bf k})$ and $\partial_{k_y}
P_{\sigma}({\bf k})$ using the symmetric finite difference as
\begin{align*}
\partial_{k_x} P_{\sigma}({\bf k}) \approx \frac{P_{\sigma,i+1,j}-P_{\sigma,i-1,j}}{2h},\\
\partial_{k_y} P_{\sigma}({\bf k}) \approx \frac{P_{\sigma,i,j+1}-P_{\sigma,i,j-1}}{2h}.
\end{align*}
Thus, in equation (\ref{eqn:chernnumb}) we simplify
\begin{eqnarray}
& & P_{\sigma}({\bf k}) \Big[\partial_{k_x} P_{\sigma}({\bf k}),
\partial_{k_y} P_{\sigma}({\bf k}) \Big] \nonumber \\ &\approx &
\frac{P_{\sigma,i,j}}{4h^2}\Big([P_{\sigma,i+1,j},P_{\sigma,i,j+1}]
+ [P_{\sigma,i,j+1},P_{\sigma,i-1,j}] \nonumber \\ &         &+
[P_{\sigma,i-1,j},P_{\sigma,i,j-1}]
+[P_{\sigma,i,j-1},P_{\sigma,i+1,j}] \Big). \nonumber
\end{eqnarray}
Note that due to periodic boundary conditions in BZ,
$P_{\sigma,L_x+1,j} \equiv P_{\sigma,1,j}$ and
$P_{\sigma,i,L_y+1}\equiv P_{\sigma,i,1}$.  Then the
Chern number is approximated as
\begin{eqnarray}
& & C_{\sigma} = \frac{1}{2\pi i}\int_{\text{B.Z.}}\text{Tr}\, \left(P_{\sigma}\,dP_{\sigma}\wedge dP_{\sigma}\right)\nonumber \\
&\approx & \frac{1}{2\pi i}\sum_{i,j=1}^N
\frac{P_{\sigma,i,j}}{4}\Big([P_{\sigma,i+1,j},P_{\sigma,i,j+1}] +
[P_{\sigma,i,j+1},P_{\sigma,i-1,j}] \nonumber
\\ & & + [P_{\sigma,i-1,j},P_{\sigma,i,j-1}] +[P_{\sigma,i,j-1},P_{\sigma,i+1,j}] \Big).
\label{eqn:projector}
\end{eqnarray}
Under such  a construction, the evaluation of the spin Chern number
might be subject to finite-size effects and  an integral Chern
number is not guaranteed. However, as we  have presented in this
paper, the approach is still useful  in characterizing topological
phase transitions which involve Chern number variations.

\end{document}